\documentclass[10pt,sigconf,letterpaper,nonacm]{acmart}

\usepackage{booktabs} 
\usepackage{tabularx, hyperref}
\usepackage[utf8]{inputenc}
\usepackage{cleveref}
\usepackage{adjustbox}
\usepackage[font=footnotesize,labelformat=simple]{subcaption}

\usepackage{xurl}
\usepackage{amsmath,amsfonts,chemarrow,balance, graphicx, subcaption, url, multirow, graphics, booktabs, colortbl, algpseudocode, xcolor}
\usepackage{xspace}
\usepackage{CJKutf8}

\usepackage{adjustbox} 
\usepackage{makecell} 

\usepackage{pifont}
\newcommand{\circleone}{\ding{202}\xspace}
\newcommand{\circletwo}{\ding{203}\xspace}
\newcommand{\circlethree}{\ding{204}\xspace}

\usepackage{tikz}
\newcommand{\fullcircle}{\tikz\draw[fill=black,draw=black] (0,0) circle (0.8ex);} 
\newcommand{\emptycircle}{\tikz\draw[fill=white,draw=black] (0,0) circle (0.8ex);} 

\usepackage{fontawesome}

\newcommand{\BfPara}[1]{\vspace{1mm}\noindent\textbf{#1.}\xspace}

\begin{document}

\date{}

\author{Mario Beluri}
\email{mabe00023@stud.uni-saarland.de} 
\affiliation{%
  \institution{Saarland University}
  \country{}
}

\author{Bhupendra Acharya}
\email{bhupendra.acharya@cispa.de}
\affiliation{%
  \institution{CISPA}
  \country{}
}

\author{Soheil Khodayari}
 \email{soheil.khodayari@cispa.de} 
\affiliation{%
 \institution{CISPA}
 \country{}
 }

\author{Giada Stivala}
\email{giada.stivala@cispa.de} 
\affiliation{%
  \institution{CISPA}
  \country{}
  }

\author{Giancarlo Pellegrino}
\email{pellegrino@cispa.de} 
\affiliation{%
  \institution{CISPA}
  \country{}
}

\author{Thorsten Holz}
\email{holz@cispa.de} 
\affiliation{%
  \institution{CISPA}
  \country{}
}

\title{Exploration of the Dynamics of Buy and Sale of Social Media Accounts}


\maketitle

\pagestyle{plain}
\pagenumbering{arabic}

\subsection*{Abstract}

There has been a rise in online platforms facilitating the buying and selling of social media accounts. While the trade of social media profiles is not inherently illegal, social media platforms view such transactions as violations of their policies. They often take action against accounts involved in the misuse of platforms for financial gain. This research conducts a comprehensive analysis of marketplaces that enable the buying and selling of social media accounts.

We investigate the economic scale of account trading across five major platforms: \emph{X}, \emph{Instagram}, \emph{Facebook}, \emph{TikTok}, and \emph{YouTube}. From February to June 2024, we identified 38,253 accounts advertising account sales across 11 online marketplaces, covering 211 distinct categories. The total value of marketed social media accounts exceeded \$64 million, with a median price of \$157 per account. Additionally, we analyzed the profiles of 11,457 visible advertised accounts, collecting their metadata and over 200,000 profile posts. By examining their engagement patterns and account creation methods, we evaluated the fraudulent activities commonly associated with these sold accounts. Our research reveals these marketplaces foster fraudulent activities such as bot farming, harvesting accounts for future fraud, and fraudulent engagement. Such practices pose significant risks to social media users, who are often targeted by fraudulent accounts resembling legitimate profiles and employing social engineering tactics. We highlight social media platform weaknesses in the ability to detect and mitigate such fraudulent accounts, thereby endangering users. Alongside this, we conducted thorough disclosures with the respective platforms and proposed actionable recommendations, including indicators to identify and track these accounts. These measures aim to enhance proactive detection and safeguard users from potential threats.
\section{Introduction}
\label{sec:introduction}

With the widespread use of social media platforms and the growing number of users, fraudsters are increasingly exploiting platforms and their users with various social engineering tricks~\cite{socialmediaScamRise1, socialmediaScamRise2, socialmediaScamRise3}. According to the Federal Trade Commission (FTC), scams originating from social media resulted in reported losses totaling \$2.7 billion between January 2021 and June 2023~\cite{socialmediafraudFTC}. The report also highlighted that fraud originating from social media accounted for higher monetary losses compared to other methods of contact.

Scammers abuse social media users through sophisticated fraudulent schemes, often organized in a complex manner, and leverage a large number of fake accounts to add layers of sophistication to their attacks~\cite{xiao2015detecting}. Beyond traditional abuses such as phishing~\cite{chhabra2011phi}, investment fraud~\cite{mirtaheri2021identifying}, and impersonation~\cite{acharya2024imitation}, scammers are found to engage in various fraudulent schemes such as clickbait and engagement farming to generate revenue~\cite{jain2021clickbait}; spreading manipulated news for political or financial gain~\cite{abdelnabi2023fact}; using deepfakes or generative AI for romance scams or pig-butchering~\cite{maras2024deconstructing}; advertising counterfeit goods~\cite{socialmediafakeproducts};  executing shipping scams~\cite{shippingscams}; fraudulent recruitment~\cite{fraudjobs}; blackmail~\cite{blackmail} and sextortion~\cite{sextortion}; and influence campaigns to manipulate public opinion or trends~\cite{influencefraud}. These fraudulent schemes have escalated in scale operations in the last years, leading to unprecedented exploitation of social media platforms and their users. 

Scammers likely find social media as an ideal platform for exploitation due to its ease of account creation compared to fake website setups. Account setup in social media profiles lacks steps such as purchasing domains and certificates or ensuring the domain does not get flagged. Social media accounts offer greater immunity compared to traditional attack vectors like email, phone, or websites. With the rise of websites that facilitate the \emph{purchase} of social media accounts, scammers gain quick and credible entry points to exploit others. These pre-established accounts often come with out-of-the-box public metrics such as followers, likes, and interactions, making them appear authentic and trustworthy to potential victims. This \emph{perceived legitimacy} enables scammers to carry out fraudulent activities with reduced suspicion. Furthermore, acquiring accounts with an existing audience allows scammers to bypass the effort of building followers organically, enabling them to scale operations rapidly. These accounts also help evade platform detection by posing as genuine users, making it more challenging for social media platforms to identify and block malicious activities.

Our research is motivated by combining the above three key perspectives: (i) the growing number of social media users, (ii) the expansion and scale of scams on social media beyond traditional attacks, and (iii) the rise of websites selling social media accounts, which allow fraudsters to bypass effort, save time, scale operations, and evade platform detection. In this paper, we analyze the marketplaces that allow buying social media accounts. Our analysis provides a comprehensive study of the marketplaces and the accounts being sold, including their public engagement metrics and scam operations. Our work addresses the critical gap in understanding what happens when such accounts are purchased and how they are subsequently misused. 

In this work, we perform a comprehensive evaluation of data collected from 11 marketplaces and over 38,000 accounts listed for sale. We analyze the processes sellers use to create accounts on these marketplaces, analyze how social media accounts are advertised and set up, and evaluate profile engagement metrics exploited for abuse. Furthermore, we quantify the impact and categorize the types of abuse associated with these accounts. Our work represents the first large-scale analysis of 11 marketplaces selling social media accounts of five platforms: \emph{X} (formerly known as Twitter), \emph{Instagram}, \emph{Facebook}, \emph{TikTok}, and \emph{YouTube}. More specifically, we identified 38,253 accounts on five social platforms which resulted in over \$64 million value for sale. We performed the tracking and comprehensive evaluation of 11,457 accounts that provided the links pointing to their respective social media platforms. Through evaluating account creation and post engagement from all five social media platforms, we shed light on how these accounts are later misused to target users. Additionally, we offer recommendations to mitigate such scams. In summary, we make following key contributions.

\begin{itemize}
    \item We conduct the first large-scale empirical study of marketplaces involved in selling social media accounts, uncovering fraudsters targeting over 210 categories as part of their scams.

	\item We provide a comprehensive evaluation of profile engagement and account creation setups, identifying the operational scale and abuse categories in which these accounts are exploited to target platforms and their users.

	\item Finally, we propose recommendations and distill insights to combat such processes, aiming to prevent the emerging threats posed by these marketplaces and sold accounts.

\end{itemize}

To foster research, we share our code~\cite{researchcode} related to marketplaces that were publicly advertised for selling. However, for data protection reasons, the data related to the study are only shared with interested academics, abused entities, or researchers upon request.

\BfPara{Ethical Consideration and Data Disclosure} Our research does not involve direct interaction with scammers or individuals, and relies solely on publicly available data. During data collection, we ensured that no engagement with human subjects occurred; our methodology was entirely passive and limited to publicly available data. Additionally, we disclosed relevant information, such as social media profiles, to all five social media platforms involved. For account-selling marketplaces, we ensured ethical compliance by refraining from bypassing CAPTCHAs, paywalls, evading automation triggers, and avoiding any direct interactions with social media profiles during automated data collection. Our findings were shared with all five social media platforms that we studied. We received positive feedback from X, expressing further interest in future collaboration.

\section{Background and Related Work}
The proliferation of social media platforms has brought transformation changes to communication, entertainment, and business. However, alongside these benefits, social media has also become a fertile ground for abuse, fraud, and malicious activity. Bad actors exploit the platforms’ vast user bases for illicit purposes such as account trading, spam, scams, and phishing, often causing harm to legitimate users and undermining trust in these platforms. Understanding the mechanisms and economic incentives behind these malicious activities is critical for designing effective countermeasures.

One growing area of concern is the emergence of fraudulent marketplaces that facilitate the trade of compromised or fake social media accounts. These accounts are used to amplify malicious campaigns, manipulate engagement metrics, or perpetrate fraud at scale. While previous studies have explored the broader cybercrime economy~\cite{stringhini2012poultry, dekoven2018following, thomas2013trafficking}, little attention has been paid to the life cycle of these marketplaces and their accounts—from their sale to their eventual involvement in abuse, which we address in this work. We provide a framework for understanding the engagement and abuse patterns of these accounts after they are traded. Our work provides a comprehensive study of identifying accounts that are solid across various marketplaces and tracking the account's engagement and abuse categories. In the following, we discuss previous works related to our study and highlight our unique nature of work addressing the research gap.

\BfPara{Cybereconomics and Fraudulent Marketplaces} 
Some of the prior work that relates to ours focused on identifying malicious services or merchants and evaluating their business model and product offering over time~\cite{stringhini2012poultry, dekoven2018following, thomas2013trafficking}. 
Stringhini et al.~\cite{stringhini2012poultry} analyzed the operations of Twitter Account Markets, which generate revenue by exploiting networks of followers, often through artificial inflation of follower counts or using compromised accounts to distribute promotional or abusive content. Similarly, DeKoven et al.~\cite{dekoven2018following} studied the for-profit services that drive traffic to manipulate the user's perception, while Thomas et al.~\cite{thomas2013trafficking} studied the role of the underground market in contributing towards abusive behavior such as scams and spams. However, none of these studies focus on identifying the accounts that are being sold and later tracking them to understand the maliciousness.

\BfPara{Detection of fake accounts} Another line of work similar to ours focused on the detection of fake accounts on social media, for example by constructing ``social profiles'' of users, allowing to detect discrepancies of the regular behavior (e.g.,~\cite{Ruan2016Profiling, egele2013compa, Cao2014Uncovering}) or by developing anomaly detection algorithms (e.g.,~\cite{traang2015evaluating, viswanath2014towards}).
Further research developed detection techniques based on the characteristics of Twitter accounts and posts (e.g.,~\cite{khalil2017detecting, cresci2015fame, stringhini2013follow, stringhini2012poultry}), on the connections between profiles (e.g.,~\cite{mehrotra2016detection}), or on the combination of multiple features (e.g.,~\cite{aggarwal2015what, xu2021deep}).
Finally, Kurt et al.~\cite{thomas2013trafficking} studied patterns in the naming and registration processes of Twitter accounts, deriving patterns allowing to detection of abusive bulk registration of profiles. Our study differs from previous work, mainly in understanding the origination of the social media accounts that are later abused in scale.

\BfPara{Spam, Scam, and Phishing} Miscreants use social media platforms to spread spammy, malicious, or scam content~\cite{Stringhini2010Detecting, Cao2014Uncovering, Acharya2024Conning}, leveraging a post's content\cite{gao2010detecting}, visual appearance\cite{stivala2020deceptive}, or the reputability of a profile\cite{chhabra2011phi, gao2010detecting}, putting legitimate users at risk. Previous works measured the number of spam tweets and URLs, finding tweets containing over 2 million distinct URLs pointing to blacklisted scams, phishing, and malware over the period of two months~\cite{grier2010spam}, and showing that most accounts spreading malicious tweets are likely compromised~\cite{grier2010spam}, although new accounts are also registered specifically with this purpose. A similar study~\cite{gao2010detecting} conducted on Facebook confirmed this phenomenon, observing how compromised accounts are used to contact victim users posting URLs leading to advertisements, phishing, and drive-by downloads. Our work complements such studies by focusing on social media and various types of scams.

\section{Evaluation Setup}
\label{sec:background}
\begin{figure}[tb]
\centering
\includegraphics[width=.48\textwidth]{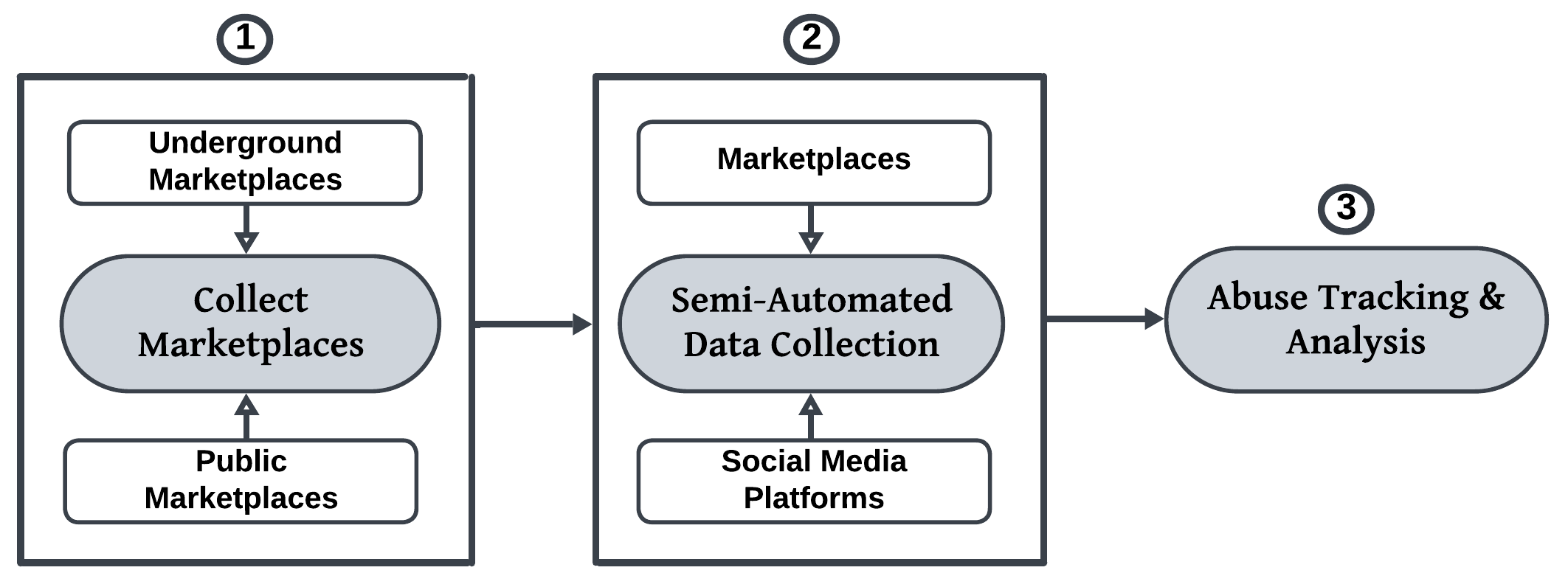}\hfill
\caption{Evaluation Setup – Our evaluation setup comprises three main modules. Initially, we collect marketplaces that sell social media accounts based on manual search (\circleone). Through semi-automation, we collect data from various marketplaces related to advertised accounts and further query social media APIs to collect data related to the advertised account (\circletwo); finally, we evaluate the collected data by analyzing the marketplace, and their affiliated social media accounts including scam tracking and abusive elements associated with such accounts (\circlethree).}
\label{fig:sys_design}
\end{figure}

In this section, we provide detailed information on evaluation setup and data collection process. Our evaluation framework consists of three main modules, as illustrated in ~\autoref{fig:sys_design}. Initially, we identify marketplaces that advertise the buying and selling of social media accounts (\circleone). Once identified, we curate these marketplaces based on the feasibility of data collection and proceed with semi-automated steps to gather advertised accounts. We then query the respective social media platforms to collect publicly available engagement and profile metadata linked to these advertised accounts (\circletwo). Finally, we track and analyze the collected marketplaces and social media accounts to uncover the mechanics and operations behind these scams (\circlethree). Below, we provide a detailed explanation of each of the three modules.

\subsection{Collect Marketplaces} 
The market for buying and selling social media accounts is divided into public and underground markets. Public markets are accessible through standard internet searches and operate with a semblance of legitimacy, often hiding behind the guise of marketing services. In contrast, underground markets are clandestine, often accessible only via specific forums or onion directories on the dark web. These underground markets operate in secrecy to avoid detection and enforcement actions.

For data collection, we initiated our investigation through Google searches and a review of previous academic papers listing account-selling websites or underground markets~\cite{motoyama11analysis, bitaab2023beyond, stringhini2013follow, dekoven2018following, lin2024malla, thomas2013trafficking, lykousas2023cynicism, li2024understanding}. This preliminary research provided a foundation, which was further expanded by tracking postings in publicly accessible forums and onion directories that list underground market sites. This dual approach ensured a comprehensive understanding of both market types. This resulted in a comprehensive list of 58 websites and nine personal contact points (emails, phone numbers, telegram handlers).
We focused on trading channels where social media account handles were publicly visible, excluding others from further automated data collection, as reported in \autoref{tab:appendix_whole_collected_data} in the Appendix. 

\subsection{Data Collection}
Our data collection relied on two primary sources: \textit{(i)} accounts advertised by sellers on various marketplaces and \textit{(ii)} for each account with a visible social media profile link, we queried the respective social media platforms to gather associated profile metadata and engagement posts. We provide further details below.

\BfPara{Public Marketplace Account Collection}
We developed a JavaScript-enabled web crawler to automatically extract account-selling offers related to popular social media platforms. These include \emph{X}, \emph{Instagram}, \emph{YouTube}, and \emph{TikTok}. The crawler is implemented using \emph{Python}, with \emph{Selenium} for browser automation and the \emph{Chrome DevTools Protocol} for fine-grained page control and interaction. 

Each marketplace may have multiple listings covering various social media platforms, such as Instagram and Twitter. For each marketplace, we manually identified the seed URLs for different listings and initialized our crawler with these URLs. Given a seed URL, the crawler employs a depth-first strategy: it visits a listing page, clicks on each offer to reach the offer webpage, and collects its details. This process continues until all offers on a listing page are covered. The crawler then moves to the next listing page and repeats the process, stopping only when no new offers or listing pages are found. For each advertised account for sale, we collect displayed information such as offer URL, title, seller information, price, payment methods, social media account handles, account properties (such as the number of likes and followers), and the offer description. 
Out of 58 trading markets, 11 contain selling offers with publicly visible social media account handles, which we focused on. In ~\autoref{table:marketplace_overview_table}, we display each marketplace, seller, and advertised account detail. In total, we collected 38,253 URLs from the 11 marketplaces, out of which 11,457 URLs display accounts linked to respective social media platforms. Among these marketplaces, 35\% (13,665/38,253) of accounts resulted from \emph{Accsmarket} as the highest number of accounts, and the lowest accounted from \emph{FameSeller} 109 accounts, lesser than 1\% of the total accounts found. We identified, 5/11 marketplaces, \emph{SocialTradia}, \emph{TooFame}, \emph{SwapSocials}, \emph{Surgegram}, and \emph{BuySocia} omitting the public display of seller's information. 

\begin{table}[tb]
    \small
    \caption{Overview of the public marketplace sellers and advertised social media accounts for sale. Among these 11 marketplaces, \emph{Accsmarket} was found to have the highest number of advertised accounts, and \emph{FameSeller} was the lowest.}
    \centering
    \begin{tabular}{lrr}
        \toprule
        \rowcolor{gray!0}
        \textbf{Public Marketplace} & \textbf{Sellers} & \textbf{Accounts} \\
        \midrule
		Accsmarket & 2,455 & 13,665 \\
		\rowcolor{gray!10}
		FameSwap & 6,617 & 8833 \\
		\rowcolor{gray!0}
		Z2U & 240 & 6,417 \\
		\rowcolor{gray!10}
		SocialTradia & - & 4,020 \\
		\rowcolor{gray!0}
		InstaSale & 251 & 1,950 \\
		\rowcolor{gray!10}
		MidMan & 304 & 1282 \\
		\rowcolor{gray!0}
		TooFame & - & 695 \\
		\rowcolor{gray!10}
		SwapSocials & - & 530 \\
		\rowcolor{gray!0}
		SurgeGram & - & 205  \\
		\rowcolor{gray!10}
		BuySocia & - & 547 \\
		\rowcolor{gray!0}
		FameSeller & 77 & 109 \\

        \midrule
        \rowcolor{gray!10} \textbf{Total} & 9,944 & 38,253 \\
		
        \bottomrule
    \end{tabular}    
    \label{table:marketplace_overview_table}
\end{table}

\BfPara{Profile Metadata Collection} Of 38,253 advertised accounts from open marketplaces, 29\% (11,457) of the accounts for sale advertised visible links pointing to their respective social media profile. For each of these social media accounts, we collected each profile's public profile metadata, including user profile names, descriptions, account creation dates, and engaging posts, from visible accounts linked to the advertised seller's marketplace page. For this, we utilized the respective API services~\cite{TwitterUserDetailAPI,TwitterTimelinesAPI,InstagramScraper,TelegramApifyScraper,TelegramTelemetrScraper,YouTubeScraper,facebookScraper} of the social media platforms. In \autoref{table:social_media_posts_and_accounts}, we present a detailed breakdown of the collected social media accounts and their corresponding posts. Our findings show that \emph{YouTube} accounts had the highest number of visible account profiles linked, account 54\% (6,271/11,457) of the total visible accounts, whereas the lowest count resulted from \emph{Facebook}, 5\% (649/11457) from our overall visible accounts.

\BfPara{Underground Forum Account Collection} We analyzed accounts sold on underground markets accessed via the Tor network. Initially, we manually inspected these markets to confirm their accessibility and available goods. Many markets referenced in related research were inaccessible due to takedowns, lack of directory listings, or timeout errors, while others were non-English, or did not sell digital goods. This narrowed our focus to four underground markets. We expanded our dataset by adding 16 more underground forums found in onion directories, which sold social media-related goods at the time of this first inspection. All inspected markets required user registration and implemented complex, site-specific, non-standard \emph{CAPTCHAs}. Additionally, navigation was restricted: attempts to access pages not linked within the current page resulted in blocks. Due to these limitations, we collected all account sale data manually. Data collection followed two criteria: \textit{(i)} browsing forum sections dedicated to accounts or social media, or \textit{(ii)} using forum search functionalities with keywords like \textit{[account/s | profile/s] [name of social media]}. In both cases, we recorded data from the first five pages of results, up to 25 postings per social media platform.

Of the 20 markets in our initial dataset, eight did not sell social media-related goods, and four offered services like likes and followers but no accounts, leaving a final dataset of eight underground markets. In the second manual inspection, we collected for each posting the URL, title, textual content, author’s username, publication date, number of replies, price, quantity sold, and a screenshot. Differences in forum models and GUIs meant that not all fields were consistently available across forums. For example, some forums did not display the date when a message was posted, or disallowed comments under the listings.

\begin{table}[tb]
    \small
    \caption{Overview of the social media data collection. In the table, we display social media accounts advertised for sale. The accounts that are listed referencing the social media platform account profile we name them as visible accounts.}
    \centering
    \begin{tabular}{lrrr}
        \toprule
        \rowcolor{gray!0}
        \textbf{Social} & \textbf{Visible} & \textbf{Visible } & \textbf{All}\\
    \textbf{Media} & \textbf{Accounts} & \textbf{Accts. Posts } & \textbf{Accounts}\\
        \midrule
			Instagram & 2,023 & 4,207 & 12,658\\
			\rowcolor{gray!10}
			YouTube & 6,271 & 3,411 & 9,087\\
                Tiktok & 1700 & 25,131 & 8,973\\
			\rowcolor{gray!10}
                Facebook & 649 & 7,407 & 4,216\\
			\rowcolor{gray!0}
                X & 814 & 165,427 & 3,319\\
        	
            \midrule
            \rowcolor{gray!10} 
            \textbf{Total} & 11,457 & 205,583 & 38,253 \\
		 
        \bottomrule
    \end{tabular}    
    \label{table:social_media_posts_and_accounts}
\end{table}

\subsection{Tracking and Analysis} The third module analyzes the data collected from marketplaces and analyzes the engaged posts from the advertised accounts. This includes aspects such as the intricacies of sellers' advertisements, public engagement with social media profiles, and abusive elements such as scam tactics and the operations of scammers targeting social media users. 

We present our findings as follows: an overview of marketplaces in ~\Cref{sec:overview_marketplaces}; profile creation and engagement analysis in ~\Cref{sec:profile_creation_and_engagement}; scam clustering and abuses in ~\Cref{sec:scam_post_analysis}; tracking and network analysis on ~\Cref{sec:scam_tracking_and_network_analysis}, efficacy and abuse control in~\Cref{sec:efficacy_and_fraud_control} and finally we provide recommendation to fight against such scam in ~\Cref{sec:recommendations}.

\section{Anatomy of Marketplaces}
\label{sec:overview_marketplaces}
We conducted a comprehensive analysis of both open and underground marketplaces involved in the buying and selling of social media accounts. Our motivation to study both types of marketplaces was to understand a broader spectrum of account trade ecosystems—ranging from visible, mainstream practices to hidden, and illicit operations. We provide detailed insights for each section below.

\subsection{Anatomy of Public Marketplaces}
\label{sec:overview_public_markerplaces}

In this section, we outline how sellers set up their profiles in public marketplaces to advertise their accounts. Specifically, we analyze into categories, account monetization, verification, descriptions, public metrics, and account pricing. We provide further detailed information as below.

\begin{figure}[tb]
\centering
\includegraphics[width=.40\textwidth]{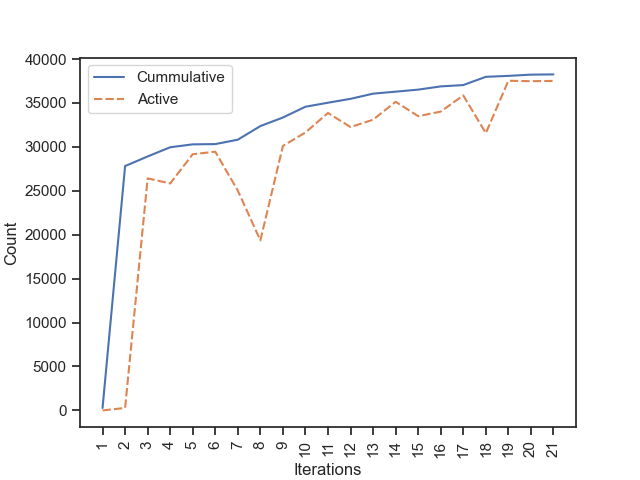} \hfill
\caption{In this graph, we present the cumulative and active listings advertised by sellers across 11 open marketplaces over our data collection iterations between Feb 2024 to Jun 2024. The decline in active listings suggests that some accounts went offline, possibly due to successful sales or the seller's decision to take them offline. At the same time, the continuous growth in cumulative listings, despite the dip in active ones, reflects the replenishment of inventory to maintain higher stock levels and meet supply and demand needs. }
\label{fig:activeness}
\end{figure}

\BfPara{Seller} We identified 9,949 sellers across 11 marketplaces. The highest number of sellers composite from \emph{FameSwap} with 6,617 sellers, while 5/11 marketplaces \emph{BuySocia}, \emph{SocialTradia}, \emph{SurgeGram}, \emph{SwapSoul}, and \emph{TooFame} lacked sufficient seller information. The median number of seller accounts was 77. Regarding seller nationality, 29,420 sellers did not disclose their country of origin, while 8,833 sellers represented 138 countries. Among these, the top five countries were the \emph{United States} (2,683 sellers), \emph{Ethiopia} (844), \emph{Pakistan} (596), the \emph{United Kingdom} (382), and \emph{Turkey} (366). In ~\autoref{fig:activeness}, we showcase the cumulative growth and activity of listings across the data collection iterations. Our observations suggest that accounts are replenished to align with supply and demand, ensuring readiness for future sales opportunities.

\BfPara{Categories Analysis} Out of 38,253 accounts, 22\% (8,775) were found to lack any categorical representation. Among the remaining 29,478 advertised accounts, 212 unique categories were identified. The top five categories were \emph{Humor/Memes} (5,056 accounts), \emph{Luxury/Motivation} (2,292), \emph{Games} (1,062), \emph{Fashion/Style} (1,678), and \emph{Reviews/How-to} (1,420). The median account size for these categories was 3.

\BfPara{Verified Accounts} Out of 38,253 accounts, we identified 185 with verified social media statuses, all of which were YouTube accounts. However, these accounts did not provide URLs linking to their respective YouTube channels. It is likely that sellers use this strategy to attract potential buyers.

\begin{table}[tb]
    \centering
    \caption{Payment methods supported by different platforms.}
    \label{tab:payment_methods}
    \footnotesize
    \centering
     \setlength{\tabcolsep}{3pt}
    
    \begin{tabular}{l|*{11}{c}}
        \textbf{Payment Methods} & 
        \makecell[tl]{\rotatebox[origin=c]{90}{Accsmarket}} & 
        \makecell[tl]{\rotatebox[origin=c]{90}{FameSwap}} & 
        \makecell[tl]{\rotatebox[origin=c]{90}{Z2U}} & 
        \makecell[tl]{\rotatebox[origin=c]{90}{SocialTradia}} & 
        \makecell[tl]{\rotatebox[origin=c]{90}{InstaSale}} & 
        \makecell[tl]{\rotatebox[origin=c]{90}{MidMan}} & 
        \makecell[tl]{\rotatebox[origin=c]{90}{TooFame}} & 
        \makecell[tl]{\rotatebox[origin=c]{90}{SwapSocials}} & 
        \makecell[tl]{\rotatebox[origin=c]{90}{SurgeGram}} & 
        \makecell[tl]{\rotatebox[origin=c]{90}{BuySocia}} & 
        \makecell[tl]{\rotatebox[origin=c]{90}{FameSeller}} \\ 
        \toprule

        \multicolumn{12}{l}{\textbf{Traditional}} \\

        \rowcolor{gray!0} \text{Visa}           &  &  & \checkmark &  &  &  &  &  & \checkmark &  &  \\ 
        \rowcolor{gray!10} \text{PayDirekt}      &  &  & \checkmark &  &  &  &  &  &  &  &  \\ 
        \rowcolor{gray!0} \text{GPay Visa}      &  &  &  &  &  & \checkmark &  &  &  &  &  \\ 
        \rowcolor{gray!10} \text{DLocal}         &  &  &  &  &  & \checkmark &  &  &  &  &  \\
        \rowcolor{gray!0} \text{Appota Visa}    &  &  &  &  &  & \checkmark &  &  &  &  &  \\

        \midrule

        \multicolumn{12}{l}{\textbf{Prepaid Vouchers}} \\
        \rowcolor{gray!10} \text{NeoSurf}        &  &  & \checkmark &  &  &  &  &  &  &  &  \\

         \midrule
        \multicolumn{12}{l}{\textbf{Crypto}} \\
        \rowcolor{gray!0} \text{BTC}            &  &  &  &  &  & \checkmark &  & \checkmark &  & \checkmark &  \\ 
        \rowcolor{gray!10} \text{ETH}            &  &  &  & \checkmark &  & \checkmark &  & \checkmark &  & \checkmark &  \\ 
        \rowcolor{gray!0} \text{LiteCoin}       &  &  &  &  &  & \checkmark &  &  &  &  &  \\ 
        \rowcolor{gray!10} \text{Tether}         &  &  &  &  &  & \checkmark &  &  &  &  &  \\ 
        \rowcolor{gray!0} \text{BNB}            &  &  &  &  &  & \checkmark &  &  &  &  &  \\ 
        \rowcolor{gray!10} \text{Matic}          &  &  &  &  &  & \checkmark &  & \checkmark &  &  &  \\ 
        \rowcolor{gray!0} \text{Dash}           &  &  &  &  &  &  &  &  &  &  &  \\

         \midrule
        \multicolumn{12}{l}{\textbf{Exchanges}} \\
        \rowcolor{gray!0} \text{Coinbase}       &  &  & \checkmark &  &  &  &  & \checkmark &  &  &  \\ 
        \rowcolor{gray!10} \text{AirWallex}      &  &  & \checkmark &  &  &  &  &  &  &  &  \\

         \midrule
        \multicolumn{12}{l}{\textbf{Digital Wallets}} \\
        
        \rowcolor{gray!0} \text{PayPal}         &  &  & \checkmark &  &  &  &  &  &  &  & \checkmark \\ 
        \rowcolor{gray!10} \text{Trustly}        &  &  & \checkmark &  &  &  &  &  &  &  &  \\ 
        \rowcolor{gray!0} \text{Skrill}         &  &  & \checkmark &  &  &  &  &  &  &  &  \\
        \rowcolor{gray!10} \text{WeChat}         &  &  & \checkmark &  &  &  &  &  &  &  &  \\ 
        \rowcolor{gray!0} \text{AliPay}         &  &  & \checkmark &  &  &  &  &  &  &  &  \\ 
         \rowcolor{gray!10} \text{Payssion}       &  &  &  &  &  & \checkmark &  &  &  &  &  \\ 

          \midrule
        
        \multicolumn{12}{l}{\textbf{Escrow-Based}} \\

        \rowcolor{gray!0} \text{Trustap}        &  &  &  &  &  & \checkmark &  & \checkmark &  &  &  \\ 
        \rowcolor{gray!10} \text{Payer}          &  &  &  &  &  & \checkmark &  &  &  &  &  \\

         \midrule
        \rowcolor{gray!0} \text{Unknown}        & \checkmark & \checkmark &  &  & \checkmark &  & \checkmark &  &  &  & \checkmark \\ 

        \bottomrule
    \end{tabular}

\end{table}

\BfPara{Account Monetization} We identified 164 accounts reporting monthly revenue generation ranging from \$1 to \$922, with a median value of \$136 and a total combined revenue of \$42,019 per month. Some sellers provided additional details about income sources and the potential benefits buyers could gain from purchasing these accounts. In total, 1,020 sellers disclosed unique income sources. The top three narratives included: \textit{(i)} generic ad-based revenue (335 sellers), \textit{(ii)} Google AdSense (73 sellers), and \textit{(iii)} video accounts with premium memberships or channel monetization (73 sellers).  Examples of these narratives are provided below:

\hfill

\emph{The account generates income by selling promotion plans to nft and crypto projects. You can sell tweets, retweets or some combos of boths. You can also sell weekly, middle or long term campaigns. A revenue-share is also a smart option. I can teach you everything to help you make income with my account.}

\hfill

\emph{You can monetise your content by selling promo videos or putting different watermarks on your Shorts videos for money.}

\hfill

\BfPara{Account Description} Out of 38,253 accounts, 63\% (24,293) included descriptions about the accounts. Through manual evaluation based on keyword analysis, we identified eight distinct strategies used in these descriptions: \textit{(i)} listings labeled as authentic (784 accounts), \textit{(ii)} listings labeled with "fresh and ready" accounts (157), \textit{(iii)} listings promoting business adaptability (122), \textit{(iv)} real user accounts with activity (116), and \textit{(v)} offers with original email included. Examples of a description are shown below:

\hfill

\emph{No shout outs have ever been done on the account. So the account is fresh and ready for whatever purposes you need – CPA, product promotion + sales, drop shipping, traffic generation, or simply you want to own an Instagram page with real and active users. Save yourself time and energy of starting a new account and growing it (which can take months). Enjoy the convenience and time saved.}

\hfill

\emph{Selling TikTok account with over 2.1 million followers and a viral video with 69 million views and 13.5 million likes. The account averages millions of views per video. This account has proven to be highly engaging and has attracted a large following. If you are interested in purchasing this account, please free to make an offer.}

\hfill

\begin{figure}[tb]
\centering
\includegraphics[width=.45\textwidth]{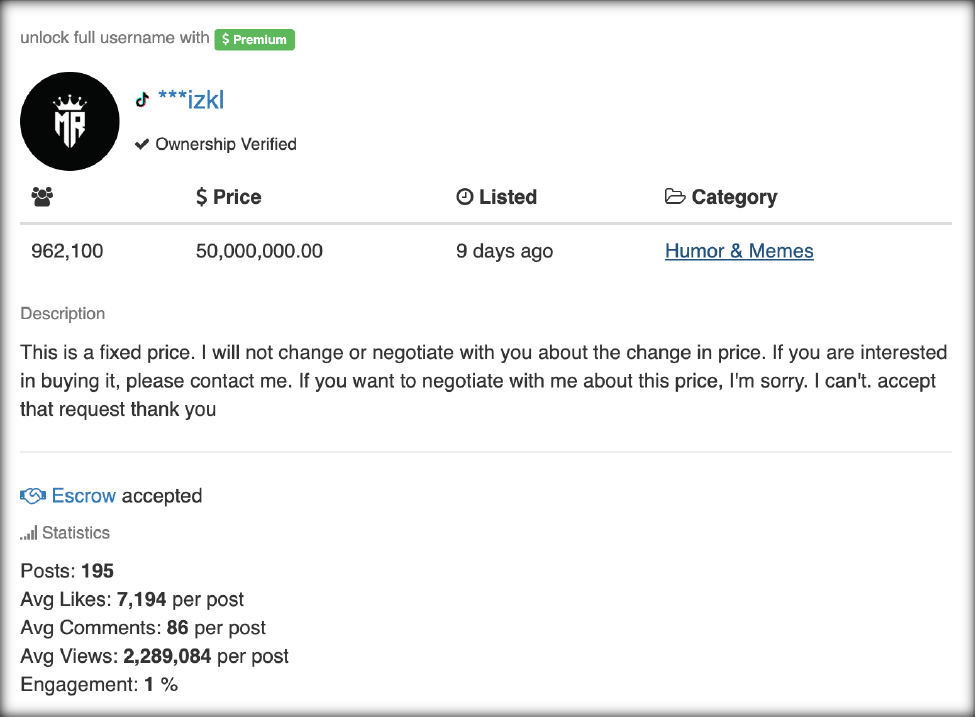} \hfill
\caption{An example of advertised seller accounts on \emph{FameSwap} marketplace with exceptionally high prices, the reasons behind such elevated prices remain unclear. The account shows the follower count close million, and the price at \$50 million.}
\label{fig:high_priced_account}
\end{figure}

\BfPara{Account Followers} Advertised accounts often share their follower counts. We found that 40\% (15,358) of accounts displayed follower information. The median follower counts for each social media platform were as follows: \emph{X} (3077), \emph{Instagram} (26,998), \emph{TikTok} (20,807), \emph{YouTube} (25,700) and \emph{Facebook} (76,050). 

\BfPara{Account Prices} The median advertised prices for social media accounts were as follows: \emph{Facebook} (\$14), \emph{X} (\$17), \emph{Instagram} (\$298), \emph{TikTok} (\$755), and \emph{YouTube} (\$759). The overall sum of all advertised account prices totaled \$64,228,836, with a median value of \$7,573,348 across the platforms. Among the five platforms, \emph{TikTok} had the highest total at \$12,760,408, while \emph{Facebook} had the lowest at \$145,937. Although the reasons behind the exceptionally high pricing of some accounts remain unclear, 345 accounts were identified with prices exceeding \$20,000. These accounts had a median price of \$45,000, a maximum of \$5,000,000, and contributed a total sum of \$38,040,411. An example of such a price account is shown in~\autoref{fig:high_priced_account}.

\BfPara{Supported Payment Methods} We analyzed the payment methods supported for buyers across 11 marketplaces. In ~\autoref{tab:payment_methods}, we present a detailed breakdown of these payment methods by the marketplace. Our findings indicate that cryptocurrency and digital wallets are preferred over traditional payment providers. This preference is likely rooted in their widespread adoption, enhanced anonymity, and reduced potential for disputes compared to traditional payment methods. In~\autoref{sec:payment_supported}, we provide additional detail in payment extraction and security implications.
\subsection{Anatomy of Underground Marketplaces}
\label{sec:underground_market}
Our investigation into underground markets for social media accounts began with an initial list of eight marketplaces: \emph{Dark Matter}~\cite{darkmatter}, \emph{Kerberos}~\cite{kerberosmarket}, \emph{Nexus}~\cite{nexusmarket}, \emph{Torzon Market}~\cite{torzonmarket}, \emph{We The North}~\cite{wethenorth}, \emph{Black Pyramid}~\cite{blackpyramid}, \emph{ARES Market}~\cite{aresmarket}, and \emph{MGM Grand}~\cite{mgmmarket}. However, at the time of our in-depth data collection, we observed that two (namely \emph{ARES Market} and \emph{MGM Grand}) did not have any account for sale, leaving six markets for analysis. These provided valuable insights into the structure and dynamics of this illicit trade. 

\BfPara{Characteristics of the Marketplaces} 
We collected a total of 65 posts from six platforms, related to four social networks. The \emph{Nexus} market offers the largest amount of accounts (37), followed by \emph{We The North} (15). The remaining four lists five or fewer accounts each, suggesting a lack of requests for this specific digital good.
Listings in these markets describe accounts for sale emphasizing characteristics like follower counts, and engagement metrics (likes and views), specifying whether they are organic or bots, whether accounts are aged, and whether they are empty or populated with content. Posts can either sell single accounts or a bulk package, sometimes creating a mismatch between the listing price and the price per account. 
The six marketplaces displayed varying levels of activity and specialization. \emph{Kerberos} had two sellers offering 51 accounts, primarily for \emph{TikTok} and \emph{X}, indicating a focus on bulk sales. The remaining markets offered one account per post. Dark Matter hosted five posts offering accounts for \emph{YouTube}, \emph{TikTok}, and \emph{X}, from three sellers. \emph{Nexus}, the most active market with 37 posts from four sellers, catered to \emph{Instagram}, \emph{X}, and \emph{TikTok}. In the \emph{Torzon} market, two sellers listed four accounts across \emph{Instagram}, \emph{TikTok}, and \emph{YouTube}, and in \emph{Black Pyramid} two sellers offered two \emph{YouTube} accounts in two posts. \emph{We The North}, with 15 posts from one single seller, exclusively targeted \emph{TikTok}, emphasizing its prominence in the underground trade. 
Among the sellers, we identified two using the same username across platforms, suggesting cross-platform operations to maximize visibility.

\BfPara{Structure and Content of Listings} 
Listings generally featured concise descriptions, with post lengths averaging between 14 and 123 words depending on the market. Sellers included contact details such as Telegram handles or website links for payment and fulfillment, as marketplaces do not handle transactions directly. Posts also frequently outlined delivery logistics, guarantees, and disclaimers about seller liability, such as for lost credentials. On the other hand, listings almost never reported the handle of the advertised product (observed only once).
Comment threads often included buyer feedback, trial requests, or "bumps" from sellers to increase visibility. Occasionally, buyers left testimonials confirming successful transactions. 

\BfPara{Patterns of Similarity Across Listings} 
A significant pattern in our analysis is the high degree of similarity observed across some posts, with word similarity ranging from 88\% to 100\%. This repetition often involves the same username (or seller) reusing identical content for multiple posts, either on a single platform or across different platforms. Interestingly, the phenomenon is even more pronounced between sellers with distinct account names, particularly within the same marketplace, and to a lesser extent across different marketplaces.

TikTok-related offerings on the Nexus market exhibit the most notable cases of textual reuse. We carried out a case-insensitive similarity analysis after removing numbers and punctuation.
For instance, we identified the same author using identical body text for two different posts (100\% similarity), seven posts from three distinct sellers with highly similar content (average similarity of 98\%), and two posts by the same seller on separate platforms, with identical text. Additionally, we found a single instance of two distinct authors posting identical text on different platforms. Altogether, 12 of the 42 posts analyzed displayed such high similarity, with all cases linked to just three authors. This consistency may suggest a coordinated effort rather than random duplication.
Similar patterns were also identified for other platforms, though less frequently, with 2 out of 13 reused posts linked to Instagram (also involving the Nexus marketplace), 1 out of 3 for Twitter, and 3 out of 7 for YouTube. 

\subsection{Comparative Summary} Our observations on underground and open marketplaces showed that \textit{(i)} underground forums are restricted via dark-web; \textit{(ii)} sellers are not very informative and often operate under pseudonyms, whereas those on open marketplaces typically disclose limited identity information; \textit{(iii)} the number of listings advertised in underground were less than 100, while the listings on open marketplaces found to be over 38K; \textit{(iv)} listings on underground markets are ultimately similar to forum posts, resulting in sparse or missing information about account details (likes, followers); \textit{(iv)} payments on underground markets were never handled by the platform but agreed upon on a different channel between buyer and seller, also via escrow methods, while payment methods on open marketplaces were rather flexible in types of payment methods; \textit{(v)} the prices on underground markets can be unclear when purchasing in bulk, or in case of private bargaining or auctions, while the open marketplace found to contain pricing detail; and furthermore we observe there are no buyer-seller mediatory transactions involved thus buyer may find unprotected during purchase at underground marketplaces. 

\medskip
For the rest of the paper, our analysis of social media profiles will be based on the public marketplaces. 
\section{Account Setup and Engagement}
\label{sec:profile_creation_and_engagement}
In the previous section, we analyzed the anatomy of marketplaces and explored how accounts are advertised for sale. In this section, we focus on understanding how the strategic preemptive tailoring of profiles aligns with market demand. Based on our findings, these accounts are meticulously crafted to target specific categories by leveraging factors such as naming conventions, descriptions, geo-locations, account setup types, and creation dates. Our analysis reveals that the preemptive tailoring of profiles aims to mimic organic profiles, drive engagement, and build a substantial subscriber base. We provide detailed observations below

\BfPara{Account Name and Description} We observe profiles frequently adopt terms and themes associated with popular industries and interests, likely to attract a wide audience or to foster trust and credibility for malicious use. This includes, for example, \textit{(i)} trendy terms such as crypto or NFTS (e.g., Crypt Hunter), \textit{(ii)} names implying expertise or status (e.g., Mr. NFT expert), \textit{(iii)} personalization appealing to specific demographics (e.g., Kajal Kumar), \textit{(iv)} profile with the adult or sensitive theme (e.g., Massage in Riyadh), and \textit{(v)} mix of unrelated names, emojis, or terms from regional and local languages (e.g.,  
\begin{CJK*}{UTF8}{goth}
まんちカビゴン
\end{CJK*}).
The account naming inclusion of financial and gaming terms indicates likely targeting of users interested in fast wealth-building or entertainment.

\BfPara{Location} We identified 3,236 profiles that listed 140 unique locations as part of their profile address, although location entries are optional on social media profiles. Among these, the top five countries represented are the \emph{US} (1,242), \emph{India} (470), \emph{Pakistan} (222), \emph{South Korea} (156), and \emph{Bangladesh} (114). This indicates that the US is the preferred location for account creation, potentially making the profiles appear more relatable and trustworthy to victims based on their origin.

\begin{table}[tb]
    \small
    \caption{Followers - In this table we present followers minimum, media, and maximum count based on the publicly marketed available social media accounts that contain visible profile URLs to respective social media platforms. We queried each social media account and obtained public metrics such as followers. This indicates that accounts for sale often harvest large numbers of followers. }
    \centering
    \begin{tabular}{lrrr}
        \toprule
        \rowcolor{gray!0}
        \multicolumn{1}{c}{\textbf{Social Media}} & \multicolumn{1}{c}{\textbf{Min}} & \multicolumn{1}{c}{\textbf{Median}}  & \multicolumn{1}{c}{\textbf{Max}}\\
        \midrule

		TikTok & 0 & 1 & 6,893  \\
		\rowcolor{gray!10}
		X  & 55 & 2,752  & 1,078,130 \\
		\rowcolor{gray!0}
		Facebook  & 115  & 27,669 & 5,239,529\\
		\rowcolor{gray!10}
		Instagram & 1032 & 8,362 & 6,288,290  \\
		\rowcolor{gray!0}
            YouTube  & 0  & 8,460 & 20,500,000\\
		\rowcolor{gray!10}
		
        \midrule
        \textbf{All} & 0 & 7,830, & 20,500,000 \\
		\bottomrule
    \end{tabular}   
    \label{table:visible_handle_followers}
\end{table}

\BfPara{Affiliated Categories} Our observation showed that social media profiles were often tagged with platform-specific categories based on their relevance. We identified 288 distinct categories associated with 1,171 accounts. The top five categories include (i) \emph{Brand and Business} (751), (ii) \emph{Entities} (349), (iii) \emph{Interests and Hobbies} (322), (iv) \emph{Digital Assets \& Crypto} (334), and (v) \emph{Events} (219). Since categories like business, interests, and assets naturally attract public engagement due to their economic relevance, such accounts are likely to be in high demand in marketplaces for purchase.

\BfPara {Account Types} Social media accounts by default are unverified and lack restrictive settings such as protected or private modes. We identified three account types across five social media platforms: \textit{(i)} business profiles marketed as entities (193), \textit{(ii)} verified accounts (669), and \textit{(iii)} accounts with controlled settings, including private (65) and protected (5) modes. This indicates that accounts for sale are predominantly unverified or standard profiles, with relatively fewer business or restricted accounts available.

\begin{figure}[tb]
\centering
\includegraphics[width=.45\textwidth]{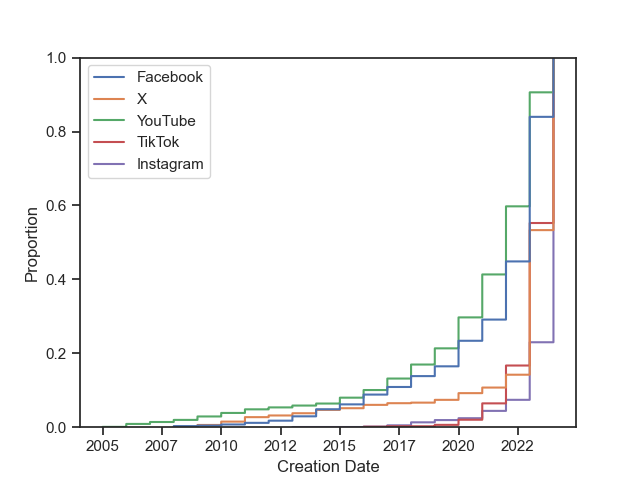} \hfill
\caption{Date of account creation - In this graph, we display visible social media account profiles from 5 social media platforms based on their date of creation date. We identify that 30\% of accounts were created before 2020, and less than 0.5\% of accounts from YouTube were created between 2006 to 2010. }
\label{fig:account_creation_date}
\end{figure}

\BfPara {Account Creation} Account creation dates provide insight into the age of social media profiles. In ~\autoref{fig:account_creation_date}, we present the CDF of account creation dates across five social media platforms. Our analysis reveals that over 70\% of accounts were created within the last 3.5 years, while less than 25\% were created between 2005 and 2020. Among the platforms analyzed, \emph{TikTok} profiles were created between 2017 and 2024, while \emph{X}, \emph{Instagram}, and \emph{Facebook} accounts date back to 2010. Notably, less than 0.5\% of \emph{YouTube} accounts were created between 2006 and 2010. This suggests that the majority of accounts advertised on these marketplaces are relatively new, with a smaller portion representing older, more established profiles.

\BfPara {Followers} Followers on social media platforms represent individuals subscribed to an account to receive notifications and view its content in their news feeds. Our analysis across five social media platforms shows that the median number of followers for accounts on sale exceeds 7,000, with the highest follower count surpassing 20 million. In ~\autoref{table:visible_handle_followers}, we present the minimum, median, and maximum follower counts for these accounts. This indicates that accounts marketed on such marketplaces are often highly engaged and likely employ engagement farming techniques to attract a substantial number of followers.

\section{Scam Post Analysis}
\label{sec:scam_post_analysis}

We perform a comprehensive evaluation to detect scam patterns given the 205K collected posts of 11.4K social media accounts (see \Cref{table:social_media_posts_and_accounts}). Our objective in analyzing these posts is to understand how fraudsters attract victims by performing various social engineering tricks. For this, we applied topic modeling techniques to group them into distinct clusters, and later performed a manual qualitative analysis of all resulting clusters to identify the scam clusters. In total, we identified six clusters performing fraudulent activities via posts. 

\BfPara{Technical Setup} Beginning with the collected posts, we focus on those based on English text, for which we rely on the CLD2 library~\cite{CLD2Library}, and remove stop words using the BERTopic library~\cite{grootendorst2022bertopic}.
Then, we extract embeddings for each post using the all-mpnet-base-v2 sentence transformer model~\cite{reimers2019sentencebert,AllMpNetBaseV2}. 
Lastly, we use HDBSCAN~\cite{mcinnes2017hdbscan} and UMAP~\cite{mcinnes2018umap} for clustering, followed by the KeyBERT model~\cite{grootendorst2020keybert} to identify potential scam posts and refine topic representations within each cluster. We then manually analyze the resulting clusters, arranged by their size,  to identify types of scam offers, and provide details on security risks. We exclude clusters that do not contain scams from our study.

\BfPara{Scam Findings} Starting with the dataset of 205K posts, we applied our methodology outlined above to automatically group the posts into 86 distinct clusters. From each cluster, we randomly selected and manually analyzed 25 sample posts to assess whether the content within the cluster was related to scams. As a result of this vetting process, we identified 16 clusters containing scam-related content, which we further categorized into six overarching scam types. 

Using this approach, we identified a total of 18.7K scam posts across over 3.7K distinct scammer accounts from five social media platforms: \emph{Facebook}, \emph{Instagram}, \emph{Tiktok}, \emph{X}, and \emph{Youtube}.  \Cref{tab:post-dataset} provides a detailed breakdown of identified scam accounts and posts, and \Cref{tab:post-analysis} presents a quick overview of the scam categories. Below, we provide an in-depth analysis of the identified scam types.

\begin{table}[tp]
	\caption{Summary of scam accounts and scam posts identified across major social media platforms. Notably, YouTube had the highest number of scam accounts, while X led in scam-related posts, highlighting significant variations in the scale of fraudulent activity across platforms.}

	\label{tab:post-dataset}
	\centering
        \footnotesize

        \begin{tabular}{l|rr}

        \toprule
        \multicolumn{1}{l|}{\textbf{Social Media}} &
        \multicolumn{1}{l}{\textbf{Scam Accounts}} &
        \multicolumn{1}{l}{\textbf{Scam Posts}} 
        \\

        \toprule
        \rowcolor{gray!0} Facebook & 512 & 3,838  \\
        \rowcolor{gray!10} Instagram & 525 & 3,271 \\
        \rowcolor{gray!0} Tiktok & 461 & 3,034 \\
        \rowcolor{gray!10} X & 610 & 6,988 \\
        \rowcolor{gray!0} Youtube & 1,661 & 1,661 \\
        
        \midrule
        \rowcolor{gray!10} \textbf{Total} & 3,769 & 18,792 \\

        \bottomrule

	\end{tabular}

\end{table}

\begin{table}[tp]
	\caption{
		Type and popularity of fraudulent offers across scammer's social media posts - This table presents six scam categories identified through post clustering. Our findings reveal that scammers frequently exploit trending topics and financial schemes, such as crypto/NFTs, while traditional scams like phishing, product fraud, adult content, and impersonation remain common.
	}
	\label{tab:post-analysis}
	\centering
    \footnotesize

    \begin{tabular}{l|rr}
    \toprule
    \multicolumn{1}{l|}{\textbf{Category}} &
    \multicolumn{1}{l}{\textbf{Accounts}} &
    \multicolumn{1}{l}{\textbf{Posts}} 
    \\
    \toprule

    \rowcolor{gray!10} \textbf{Financial Scams} & 2,649 & 8,903 \\
    \hspace{1em}- Crypto Scams & 2,352 & 8,218 \\
    \hspace{1em}- NFT and Giveaway Scams & 163 & 389 \\
    \hspace{1em}- Financial Consulting & 81 & 133 \\
    \hspace{1em}- Emotional Exploitation (Charity) & 53 & 163 \\

    \midrule
    \rowcolor{gray!10} \textbf{Phishing} & 933 & 2,293 \\
    \hspace{1em}- Through Popular Content/Challenges/Trends & 725 & 1,749 \\
    \hspace{1em}- Through Chat Communication & 208 & 544 \\
    \midrule
    \rowcolor{gray!10} \textbf{Product/Service Fraud} & 701 & 2,009 \\
    \hspace{1em}- Product Promotion Scams & 296 & 739 \\
    \hspace{1em}- Fake Travel Deals & 131 & 357 \\
    \hspace{1em}- Vehicle Sale/Rental Fraud & 101 & 279 \\
    \hspace{1em}- Sports Betting and Merchandise Scams & 129 & 451 \\
    \hspace{1em}- Fake Education-related Offers & 44 & 183 \\
    \midrule
    \rowcolor{gray!10} \textbf{Adult Content} & 244 & 466 \\
    \hspace{1em}- Provocative and Catphishing Lures & 244 & 466 \\
    \midrule
    \rowcolor{gray!10}\textbf{Impersonation} & 188 & 392 \\
    \hspace{1em}- Public Figures & 53 & 133 \\
    \hspace{1em}- Fake Tech Support & 135 & 259 \\    
    \midrule
    \rowcolor{gray!10}\textbf{Engagement Bait} & 2,300 & 4,597 \\
    \hspace{1em}- Like/Follow/Subscribe Requests & 1,509 & 2,999 \\
    \hspace{1em}- Greetings and Motivational Phrases & 791 & 1,598 \\

    \bottomrule
    \end{tabular}

\end{table}

\textbf{Financial Scams.}  
Financial scams are one of the most pervasive forms of fraudulent activity on social media, characterized by their focus on exploiting users' financial interests or vulnerabilities. These scams are perpetrated by 2,649 accounts producing 8,903 posts. A major subcategory is crypto scams, which involve promises of high returns on cryptocurrency investments, fake trading platforms, and fraudulent initial coin offerings. These scams leverage the rising popularity of digital assets to deceive users and account for 2,352 accounts and 8,218 posts. Similarly, NFT and giveaway scams capitalize on the emerging non-fungible token market by promoting fake NFT projects or false giveaways, engaging 163 accounts and 389 posts. Financial consulting scams target users seeking financial advice, with scammers impersonating consultants to extract sensitive information or money; this subcategory is responsible for 81 accounts and 133 posts. Finally, emotional exploitation scams, such as fake charity campaigns, manipulate users' goodwill by soliciting donations for fabricated causes, with 53 accounts and 163 posts contributing to this deceitful practice.

\textbf{Engagement Bait.}  
Engagement bait scams exploit users' desire for connection and social media algorithms that reward interactions. These scams involve 2,300 accounts generating 4,597 posts designed to maximize user engagement under false pretenses. Like/follow/subscribe requests, the most common type, are generated by 1,509 accounts through 2,999 posts. These requests often promise rewards or exclusive content in return for likes or follows but deliver nothing of value. Similarly, greetings and motivational phrases—posted by 791 accounts through 1,598 posts—capitalize on users' emotional responses to generic but engaging content. While appearing harmless, these tactics often serve as precursors to more deceptive practices by increasing scammers' visibility and reach.

\textbf{Phishing Scams.}  
Phishing scams are highly deceptive and aim to extract sensitive personal information such as login credentials, financial data, or identification details. These scams involve 933 accounts across 2,293 posts. One variant, phishing through popular content, challenges, or trends, mimics viral posts to lure users into clicking malicious links, with 725 accounts producing 1,749 posts. Another common form, phishing through chat communication, involves scammers directly messaging users while posing as trusted entities, accounting for 208 accounts and 544 posts. These scams exploit users' trust and curiosity, often leading to compromised accounts or financial losses.

\textbf{Product/Service Fraud.}  
Product and service fraud involves the false advertising of goods or services that do not exist, luring users with appealing offers. This category comprises 701 accounts generating 2,009 posts. Service and product promotion scams, executed by 296 accounts through 739 posts, mislead users with fake products, often using urgency to compel immediate purchases. Fake travel deals target vacationers with unrealistically cheap travel packages, involving 131 accounts and 357 posts. Vehicle sale/rental fraud, often associated with nonexistent cars or rentals, is perpetuated by 101 accounts through 279 posts. Additionally, sports betting and merchandise scams, conducted by 129 accounts through 451 posts, exploit sports fans with promises of exclusive merchandise or fixed betting outcomes.

\textbf{Adult Content Scams.}  
Adult content scams exploit the intimate nature of social media interactions to deceive users, often involving provocative imagery or fabricated romantic advances. These scams are carried out by 244 accounts across 466 posts. A typical scheme involves catfishing, where scammers pretend to be romantic interests to extract money, gifts, or sensitive information from their targets. These scams prey on users' emotions and can escalate into extortion or identity theft.

\textbf{Impersonation.}  
Impersonation scams rely on mimicking trusted entities, such as public figures or technical support services, to deceive users. This category includes 188 accounts generating 392 posts. Public figure impersonation, carried out by 53 accounts through 133 posts, involves scammers posing as celebrities or influencers to promote fake products or investment schemes. Similarly, fake tech support scams, conducted by 135 accounts through 259 posts, impersonate legitimate support agents to trick users into granting remote access to their devices or paying for unnecessary services. These scams exploit trust and authority to gain victims' compliance.

\begin{table*}[tb]
    \small
    \caption{Network Cluster Detail - In this table we provide the network analysis of social media that contain shared attributes such as name, description, biography, email, phone,e or website. Based on their profile data analysis, we cluster the accounts by these attributes and present the clustering evaluation. Our results highlight that a single cluster from \emph{Instagram} consists of as many as 46 social media accounts linked, whereas the smallest number of clusters consists of \emph{TikTok}.}
    \centering
    \begin{tabular}{lrrrrrrrr}
        \toprule
        \rowcolor{gray!0}
        \multicolumn{1}{c}{\textbf{Social Media}} & \multicolumn{1}{c}{\textbf{Cluster Attributes}} & \multicolumn{1}{c}{\textbf{Min}}  & \multicolumn{1}{c}{\textbf{Max}} & \multicolumn{1}{c}{\textbf{Median}} & \multicolumn{1}{c}{\textbf{Clusters}} & \multicolumn{1}{c}{\textbf{Cluster Accts.}} & \multicolumn{1}{c}{\textbf{Singleton}} & \multicolumn{1}{c}{\textbf{Overall Cluster Acts.}}\\
        \midrule
        TikTok & Description & 2 & 22 & 4 & 3 & 26 & 1,674 & 1.5\%\\
        \rowcolor{gray!10}
        YouTube & Name & 2 & 3 & 2 & 97 & 195 & 6,076 &  3.1\%\\
        \rowcolor{gray!0}
        Instagram & Biography & 2 & 46 & 2 & 31 & 152 & 1,871 & 7.5\%\\
        \rowcolor{gray!10}
        Facebook & Email/Phone/Website & 2 & 4 & 2 & 37 & 81 & 568 & 12.48\%\\
        \rowcolor{gray!0}
        X & Name/Description & 2 & 7 & 2 & 35 & 89 & 725 & 19.93\% \\
        \rowcolor{gray!10}
        \midrule
        \textbf{All} & - & 2 & 46 & 2 & 203 & 543 & 10,914 & 4.7\%\\
		\bottomrule
    \end{tabular}   
    \label{table:networ_analysis}
\end{table*}

\section{Tracking and Network Analysis}
\label{sec:scam_tracking_and_network_analysis}

Our network analysis of visible profiles analyzes how account formations are linked to various other social media profiles enabling us to understand the scale of the operations. We provide network evaluation below.

\BfPara{Cluster Formation} To identify cluster formations, we selected profile metadata attributes such as names, descriptions, email addresses, websites, and phone numbers. Using these attributes, we automated the clustering process to group accounts from each social media platform into buckets containing at least two or more unique UserIDs. Accounts without matching attributes across multiple profiles were categorized as singletons. After the automated clustering, for each cluster of the cluster, we perform a manual inspection to validate the legitimacy of the groupings based on these attributes. Our findings are summarized below.

\BfPara{Findings} Our findings indicate that fewer than 5\% of accounts were part of coordinated clustered campaigns. The remaining 95\% of accounts showed no significant correlation with other social media profiles based on their visible profile metadata.  In ~\autoref{table:networ_analysis}, we detail the clustering results for each social media platform, including cluster attributes, cluster sizes (minimum, median, and maximum), the total number of clusters identified, number of cluster accounts, singleton, and overall cluster accounts percentage from the dataset of each social media profiles from their respective platforms. Across the five social media platforms, a total of 203 clusters were identified, with the highest number of cluster composite from \emph{YouTube}, and the lowest number of clusters from \emph{TikTok}. Our observation showed that one of the clusters from \emph{Instagram} consists of 46 social media accounts. The median and minimum cluster size across all platforms was 2, while the total median number of clusters identified across the five social media platforms was 35, containing a median of 89 accounts per cluster.

\begin{figure}[t]
    \centering
    \caption{Three examples of the profile descriptions of advertised social media accounts.} 
    \includegraphics[width=0.95\columnwidth]{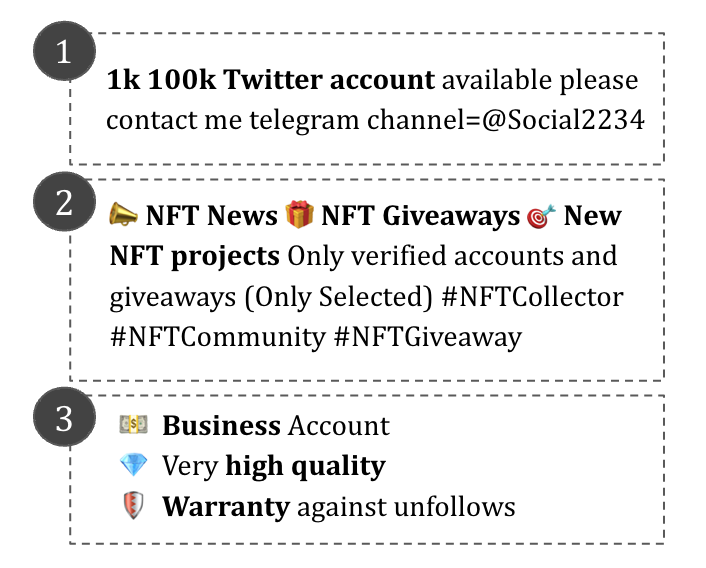}
    \label{fig:profile-example}
\end{figure}

We provide three illustrative examples of clustering based on the profile descriptions of advertised social media accounts in \Cref{fig:profile-example}.
The first example illustrates the seller harvesting 1K accounts each of those accounts having 100K X (Twitter) followers and asking users to communicate via an external communication channel (Telegram), indicating a covert and significant scale of operations designed to engage victims privately. In the second example below, an account advertises free giveaways related to NFTs, which are used as bait to lure users into scams under the guise of community engagement. The third example shows an account targeting businesses or entities, offering high-quality profiles to attract buyers in the guise of established business or promotional purposes. Thus, these show a diversity of operations, ranging from large-scale scams to targeted strategies for monetizing social media accounts, and tactics employed by sellers beyond the originated marketplaces.

\section{Efficacy and Abuse Control}
\label{sec:efficacy_and_fraud_control}

In this section, we analyze social media accounts that were actioned upon by platforms and evaluate the efficacy of blocking such accounts. 

\BfPara{Detection Overview}
We analyzed the active status of 11,457 social media profiles using API responses from the respective platforms. These responses provided explanations for account actions, such as accounts on \emph{X} being labeled as either \emph{Forbidden} or \emph{Not Found}. The \emph{Forbidden} status indicates that the account was banned due to policy violations, while \emph{Not Found} suggests that the account owner either changed their UserID or voluntarily deleted the account. On \emph{Instagram}, the status appears as \emph{Page Not Found}, while \emph{TikTok}, \emph{YouTube}, and \emph{Facebook} display messages like \emph{Profile/channel does not exist}. We suspect that accounts labeled as \emph{Not Found} or \emph{Does not exist} are likely associated with scammer or abuse profiles. Anecdotally, accounts either go offline intentionally after successfully executing scams or are taken down by the platform for violating policies during the operation of scams. We classify both scenarios under the efficacy of social media platforms in addressing and deactivating such accounts conservatively.  

\BfPara{Findings}
Out of the 11,457 accounts analyzed, the overall efficacy of social media platforms in blocking these accounts was 19.71\% (2,259 accounts). A detailed breakdown of inactive accounts and their percentages across platforms is provided in ~\autoref{table:detection_effifacy}. Among the five platforms, \emph{TikTok} and \emph{Instagram} demonstrated the highest detection efficacy at 48\%, whereas \emph{YouTube} and \emph{Facebook} showed the lowest efficacy at just 5\%. Our analysis revealed that blocked accounts frequently featured names associated with trends like \emph{crypto}, \emph{NFTs}, \emph{beauty}, \emph{luxury}, \emph{animals}, or miscellaneous word combinations. This suggests that detection efforts are largely focused on accounts leveraging popular or trending topics. Although \emph{TikTok} and \emph{Instagram} exhibited relatively higher blocking efficacy, given more than 70\% of overall visible accounts were created within the last 3.5 years---this period is substantial for scammers to cause significant harm or abuse to online users and platforms. Therefore, while platforms already shown taking proactive detection efforts on these accounts show some promise, the overall efficacy still highlights a concerning gap in addressing and preventing such threats effectively.

\begin{table}[tb]
    \small
    \caption{Detection Efficacy - In this table we present, the blocking effectiveness of social media platforms that were advertised for sale in open marketplaces. Our observations showed that \emph{TikTok} and \emph{Instagram} had overall 50\% of the blocking while \emph{X}, \emph{Facebook}, and \emph{YouTube} blocking lower than 20\% of the advertised accounts.}
    \centering
    \begin{tabular}{lrrr}
        \toprule
        \rowcolor{gray!0}
        \multicolumn{1}{c}{\textbf{Social}} & \multicolumn{1}{c}{\textbf{Visible}} & \multicolumn{1}{c}{\textbf{Inactive}}  & \multicolumn{1}{c}{\textbf{Blocking}}\\
        \multicolumn{1}{c}{\textbf{Media}} & \multicolumn{1}{c}{\textbf{Accounts}} & \multicolumn{1}{c}{\textbf{Accounts}}  & \multicolumn{1}{c}{\textbf{Efficacy}}\\
        \midrule

        YouTube & 6,271 & 315 & 5.02 \\
        \rowcolor{gray!10}
        Facebook & 649 & 37 & 5.70 \\
        \rowcolor{gray!0}
        X & 814 & 152 & 18.67 \\
        \rowcolor{gray!10}
        Instagram & 2,023 & 939 & 46.41 \\
        \rowcolor{gray!0}
        TikTok & 1,700 & 816 & 48 \\
        \rowcolor{gray!10}
        
        \midrule
        \textbf{All} & 11,457 & 2,259 & 19.71 \\
		\bottomrule
    \end{tabular}   
    \label{table:detection_effifacy}
\end{table}
\section{Recommendations}
\label{sec:recommendations}

Our study shows that accounts that are advertised for selling at these marketplaces undergo preemptive tailoring for future fraud and abuses. Thus the ecosystem of buying and selling social media profiles fosters cybercriminals to operate at scale, making it easy to obtain accounts to launch various cybercrime activities. Throughout these processes, various platforms (e.g., social media, payment vendors) and users are exploited. With that, we would like to provide recommendations for the three parties below. 

\BfPara{Social Media Platforms} We recommend social media platforms apply stricter and multi-level authenticity that discourages trading of accounts. This includes but is not limited to \textit{(i)} monitoring referral headers that are directed from marketplaces that buy and sell social media profiles, and \textit{(ii)} performing behavioral monitoring of accounts such as rapid follower growth, change of location, or IP addresses that may indicate a likelihood of engagement or account farming. Additionally, we encourage social media platforms to run public awareness campaigns highlighting the risks of account trading, which may involve compromised or illicitly obtained accounts, and to communicate the consequences of platform penalties.

\BfPara{Payment and Transaction Monitor} We recommend that payment services such as \emph{PayPal}, \emph{cryptocurrency exchanges}, \emph{wallet providers}, and similar vendors implement robust fraud detection systems to flag transactions linked to the trading of social media accounts. For example, during account verification or onboarding for payment services, a thorough analysis should be conducted to determine the intended use of the service. Similarly, payment platforms should monitor and flag addresses associated with marketplaces facilitating account sales. Establishing strict paywall transaction monitoring and reporting mechanisms would enhance the detection and prevention of fraudulent activities in this context.

\BfPara{Law Enforcement and Policy Makers} Currently, the trading of social media profiles operates in a grey area. While social media platforms view such activities as violations of their terms and conditions, such violations result in account bans, which are not explicitly illegal under current laws. This lack of regulation creates a gap in both oversight of social media account trading and consumer protection. We recommend that law enforcement agencies and policymakers explicitly ban the sale of social media profiles by incorporating clear prohibitions in legal frameworks. This is particularly critical as purchased accounts are often misused for malicious purposes. Collaborative efforts with social media companies and \emph{DNS sinkholes} should be enforced to identify and take down domains associated with marketplaces facilitating account sales. Additionally, we propose establishing robust consumer protection measures. This should include penalties for individuals or organizations found engaging in the buying or selling of social media accounts, especially in cases where such practices are likely in the future, use for exploit or defraud others.
\section{Lessons Learned}

In this section, we summarize the main findings of our study and discuss their wider implications. 

\BfPara{The Hidden Scale and Economics of Account Sales}
This paper provides the first large-scale empirical analysis of 38K social media accounts listed for sale, revealing a total market value exceeding \$64M, with median prices differing across platforms (e.g., Instagram: \$298, TikTok: \$755), providing key insights into the economic drivers of this illicit market.

\BfPara{Old Accounts, New Tricks: Creation Patterns as Fraud Tools}
We provide a novel timeline of account creation, revealing that 30\% of sold accounts were created pre-2020, leveraging their longevity to evade detection.
Conversely, accounts created in the past 3.5 years still dominate scam activity ($\sim$70\%), suggesting that scammers quickly adapt to platform changes and user trends.

\BfPara{Playbooks of Deceit: Fraud Strategies in Marketplaces}
Through analysis of both public and underground marketplaces, we identify coordinated fraud strategies, including high textual similarity (up to 100\%) across scam listings, indicating shared playbooks among fraud networks.

\BfPara{The Anatomy of Scams: Types and Tactics}
We categorize 18.7K scam posts into six distinct types, including financial scams, phishing, and impersonation. This clustering provides actionable insights into how fraudsters operate across platforms. Fraudulent accounts target specific categories (e.g., crypto, gaming, luxury) with tailored narratives to exploit niche communities, demonstrating high levels of operational precision.

\BfPara{Engagement Metrics Boost Fraudulent Credibility}
By analyzing engagement metrics from 11,457 accounts, we demonstrate how these metrics are exploited to enhance the perceived legitimacy of fraudulent accounts.
We observed that accounts are pre-configured with characteristics such as high follower counts and strategic descriptions to enhance their appeal before sale.

\BfPara{Profiling Seller Activity Across Platforms}
We identify patterns in seller activity, including cross-marketplace operations, and show how sellers replenish listings to align with supply-demand dynamics. Cross-platform activities, including identical seller profiles on the dark web and public marketplaces, highlight a merging of traditionally separate fraud ecosystems.

\BfPara{Social Media Detection Gaps}
Our results show that, despite platform efforts, only 19.7\% of identified fraudulent accounts were actioned upon by social media platforms, underscoring the critical need for enhanced detection methodologies.
\section{Concluding Remarks}

In conclusion, the rise of online marketplaces for trading social media accounts presents significant risks to platform integrity and user safety. While not inherently illegal, these transactions violate the policies of platforms like X, Instagram, Facebook, TikTok, and YouTube, and fuel fraudulent activities. Our analysis, conducted from February to June 2024, identified 38,253 accounts advertised for sale across 11 marketplaces and 211 distinct categories, representing a total value exceeding \$64 million, with a median price of \$120 per account.
We examined 11,457 visible advertised accounts and collected metadata along with over 200K associated posts. This data revealed fraudulent practices such as bot farming, account harvesting for future scams, and deceptive engagement manipulation. These fraudulent accounts often impersonate legitimate profiles, leveraging social engineering tactics to exploit unsuspecting users. Platforms currently face challenges in detecting and mitigating these threats, leaving users vulnerable to attacks. To address these issues, we provided detailed disclosures to the respective platforms and proposed practical recommendations including indicators to identify and track fraudulent accounts given the scam patterns and tactics we discovered in our research.
\section*{Acknowledgements}
The authors gratefully acknowledge funding from the German Federal Ministry of Education and Research (BMBF grant 16KIS1900 ``UbiTrans''). The authors used GPT-4 and Grammarly to revise the text of most sections to correct typos, grammatical errors, and awkward phrasing.

{\footnotesize \bibliographystyle{acm}
\bibliography{bibliography}}

\begin{thebibliography}{10}

\bibitem{aresmarket}
Ares market.
\newblock \url{https://sn2sfdqay6cxztroslaxa36covrhoowe6a5xug6wlm6ek7nmeiujgvad.link/}.

\bibitem{blackpyramid}
Black pyramid.
\newblock \url{http://blackpyoc3gbnrlvxqvvytd3kxqj7pd226i2gvfyhysj24ne2snkmnyd.onion/}.

\bibitem{darkmatter}
Dark matter.
\newblock \url{http://darkmmro6j5xekpe7jje74maidkkkkw265nngjqxrv4ik7v3aiwdbtad.onion/}.

\bibitem{kerberosmarket}
Kerberos market.
\newblock \url{http://kerberqtg7xpofsc3w47nvjd52sys6hqdejk3h7fz6kbqhyqrds3xgqd.onion/}.

\bibitem{mgmmarket}
Mgm market.
\newblock \url{https://mgmsanjqxo4svh35yqkxxe5r54z2xc5tjf6r3ichxd3m2rwcgabf44ad.xyz/}.

\bibitem{nexusmarket}
Nexus market.
\newblock \url{http://nexusabcdkq4pdlubs6wk6ad7pobuupzoomoxi6p7l32ci4vjtb2z7yd.onion/}.

\bibitem{AllMpNetBaseV2}
{Sentence Transformer all-mpnet-base-v2}.
\newblock \url{https://huggingface.co/sentence-transformers/all-mpnet-base-v2}.

\bibitem{torzonmarket}
Torzon market.
\newblock \url{http://sglgj2fytneccvyn6n4u3pacj4zhdhscfoptnhxxes3uvljmontru2yd.onion/}.

\bibitem{wethenorth}
We the north.
\newblock \url{http://hn2paw7zaahbikbejiv6h22zwtijlam65y2c77xj2ypbilm2xs4bnbid.onion/}.

\bibitem{abdelnabi2023fact}
{\sc Abdelnabi, S., and Fritz, M.}
\newblock $\{$Fact-Saboteurs$\}$: A taxonomy of evidence manipulation attacks against $\{$Fact-Verification$\}$ systems.
\newblock In {\em USENIX Security\/} (2023).

\bibitem{acharya2024imitation}
{\sc Acharya, B., Lazzaro, D., L{\'o}pez-Morales, E., Oest, A., Saad, M., Cin{\`a}, A.~E., Sch{\"o}nherr, L., and Holz, T.}
\newblock The imitation game: Exploring brand impersonation attacks on social media platforms.
\newblock In {\em USENIX Security\/} (2024).

\bibitem{Acharya2024Conning}
{\sc Acharya, B., Saad, M., Cinà, A.~E., Schönherr, L., Nguyen, H.~D., Oest, A., Vadrevu, P., and Holz, T.}
\newblock Conning the crypto conman: End-to-end analysis of cryptocurrency-based technical support scams.
\newblock In {\em IEEE Symposium on Security and Privacy (IEEE S\&P)\/} (2024).

\bibitem{aggarwal2015what}
{\sc Aggarwal, A., and Kumaraguru, P.}
\newblock What they do in shadows: Twitter underground follower market.
\newblock In {\em Annual Conference on Privacy, Security and Trust (PST)\/} (2015).

\bibitem{InstagramScraper}
{\sc Apify}.
\newblock Apify instagram scraper api.
\newblock \url{https://apify.com/apify/instagram-scraper}, 2024.

\bibitem{facebookScraper}
{\sc Apify}.
\newblock Facebook scraper.
\newblock \url{https://apify.com/streamers/facebook-scraper}, 2024.

\bibitem{YouTubeScraper}
{\sc Apify}.
\newblock Youtube scraper.
\newblock \url{https://apify.com/streamers/youtube-scraper}, 2024.

\bibitem{bitaab2023beyond}
{\sc Bitaab, M., Cho, H., Oest, A., Lyu, Z., Wang, W., Abraham, J., Wang, R., Bao, T., Shoshitaishvili, Y., and Doup{\'e}, A.}
\newblock Beyond phish: Toward detecting fraudulent e-commerce websites at scale.
\newblock In {\em 2023 IEEE Symposium on Security and Privacy (IEEE S\&P)\/} (2023).

\bibitem{CLD2Library}
{\sc Bowyer, G.}
\newblock {CLD2-CFFI – Python (CFFI) Bindings for Compact Language Detector 2}.
\newblock \url{https://github.com/GregBowyer/cld2-cff}, 2016.

\bibitem{Cao2014Uncovering}
{\sc Cao, Q., Yang, X., Yu, J., and Palow, C.}
\newblock Uncovering large groups of active malicious accounts in online social networks.
\newblock In {\em ACM SIGSAC Conference on Computer and Communications Security (CCS)\/} (2014).

\bibitem{chhabra2011phi}
{\sc Chhabra, S., Aggarwal, A., Benevenuto, F., and Kumaraguru, P.}
\newblock Phi.sh\$ocial: the phishing landscape through short urls.
\newblock In {\em Electronic Messaging, Anti-Abuse and Spam Conference (EMASC)\/} (2011).

\bibitem{cresci2015fame}
{\sc Cresci, S., {Di Pietro}, R., Petrocchi, M., Spognardi, A., and Tesconi, M.}
\newblock Fame for sale: Efficient detection of fake twitter followers.
\newblock {\em Decision Support Systems\/} (2015).

\bibitem{dekoven2018following}
{\sc DeKoven, L.~F., Pottinger, T., Savage, S., Voelker, G.~M., and Leontiadis, N.}
\newblock Following their footsteps: Characterizing account automation abuse and defenses.
\newblock In {\em ACM SIGCOMM Conference on Internet Measurement Conference (IMC)\/} (2018).

\bibitem{egele2013compa}
{\sc Egele, M., Stringhini, G., Kruegel, C., and Vigna, G.}
\newblock Compa: Detecting compromised accounts on social networks.
\newblock In {\em Network and Distributed System Security (NDSS)\/} (2013).

\bibitem{gao2010detecting}
{\sc Gao, H., Hu, J., Wilson, C., Li, Z., Chen, Y., and Zhao, B.~Y.}
\newblock Detecting and characterizing social spam campaigns.
\newblock In {\em ACM SIGCOMM conference on Internet measurement (IMC)\/} (2010).

\bibitem{grier2010spam}
{\sc Grier, C., Thomas, K., Paxson, V., and Zhang, M.}
\newblock @ spam: the underground on 140 characters or less.
\newblock In {\em ACM conference on Computer and communications security (CCS)\/} (2010).

\bibitem{grootendorst2020keybert}
{\sc Grootendorst, M.}
\newblock {KeyBERT: Minimal Keyword Extraction with BERT}.
\newblock \url{https://doi.org/10.5281/zenodo.4461265}, 2020.

\bibitem{grootendorst2022bertopic}
{\sc Grootendorst, M.}
\newblock {BERTopic: Neural topic modeling with a class-based TF-IDF procedure}.
\newblock {\em arXiv arXiv:2203.05794\/} (2022).

\bibitem{socialmediaScamRise1}
{\sc IDRC}.
\newblock Social media scams are on the rise as more people use the platforms to connect.
\newblock \url{https://www.idtheftcenter.org/post/social-media-scams-are-on-the-rise-as-more-people-use-the-platforms-to-connect/}, 2020.

\bibitem{jain2021clickbait}
{\sc Jain, M., Mowar, P., Goel, R., and Vishwakarma, D.~K.}
\newblock Clickbait in social media: detection and analysis of the bait.
\newblock In {\em Information Sciences and Systems (CISS)\/} (2021).

\bibitem{socialmediaScamRise3}
{\sc Jr., T.~H.}
\newblock Social media scams: Stunning statistics and tips to protect yourself.
\newblock \url{https://www.cnbc.com/2023/10/12/americans-lose-billions-to-social-media-scams-red-flags-to-spot.html}, 2023.

\bibitem{khalil2017detecting}
{\sc Khalil, A., Hajjdiab, H., and Al-Qirim, N.}
\newblock Detecting fake followers in twitter: A machine learning approach.
\newblock {\em International Journal of Machine Learning and Computing\/} (2017).

\bibitem{li2024understanding}
{\sc Li, Z., and Liao, X.}
\newblock Understanding and analyzing appraisal systems in the underground marketplaces.
\newblock In {\em Network and Distributed System Security (NDSS)\/} (2024).

\bibitem{lin2024malla}
{\sc Lin, Z., Cui, J., Liao, X., and Wang, X.}
\newblock Malla: Demystifying real-world large language model integrated malicious services.
\newblock {\em arXiv arXiv:2401.03315\/} (2024).

\bibitem{lykousas2023cynicism}
{\sc Lykousas, N., Koutsokostas, V., Casino, F., and Patsakis, C.}
\newblock The cynicism of modern cybercrime: Automating the analysis of surface web marketplaces.
\newblock In {\em IEEE International Conference on Service-Oriented System Engineering (SOSE)\/} (2023).

\bibitem{maras2024deconstructing}
{\sc Maras, M.-H., and Ives, E.~R.}
\newblock Deconstructing a form of hybrid investment fraud: Examining ‘pig butchering’in the united states.
\newblock {\em Journal of Economic Criminology\/} (2024).

\bibitem{mcinnes2017hdbscan}
{\sc McInnes, L., Healy, J., and Astels, S.}
\newblock {HDBSCAN: Hierarchical Density Based Clustering}.
\newblock {\em Journal of Open Source Software\/} (2017).

\bibitem{mcinnes2018umap}
{\sc McInnes, L., Healy, J., and Melville, J.}
\newblock {UMAP: Uniform Manifold Approximation and Projection for Dimension Reduction}.
\newblock {\em arXiv arXiv:1802.03426\/} (2018).

\bibitem{mehrotra2016detection}
{\sc Mehrotra, A., Sarreddy, M., and Singh, S.}
\newblock International conference on contemporary computing and informatics (ic3i).
\newblock In {\em 2016 2nd International Conference on Contemporary Computing and Informatics (IC3I)\/} (2016).

\bibitem{TelegramApifyScraper}
{\sc Milevski, D.}
\newblock Apify telegram scraper api.
\newblock \url{https://apify.com/danielmilevski9/telegram-channel-scraper}, 2024.

\bibitem{TelegramTelemetrScraper}
{\sc Milevski, D.}
\newblock Telemetrio telegram scraper api.
\newblock \url{https://telemetr.io/}, 2024.

\bibitem{blackmail}
{\sc Milmo, D.}
\newblock Sharp rise in blackmail of children asked to share explicit images.
\newblock \url{https://www.theguardian.com/society/2023/may/12/sharp-rise-in-blackmail-of-children-asked-to-share-explicit-images}, 2023.

\bibitem{mirtaheri2021identifying}
{\sc Mirtaheri, M., Abu-El-Haija, S., Morstatter, F., Ver~Steeg, G., and Galstyan, A.}
\newblock Identifying and analyzing cryptocurrency manipulations in social media.
\newblock {\em IEEE Transactions on Computational Social Systems (IEEE TCSS)\/} (2021).

\bibitem{motoyama11analysis}
{\sc Motoyama, M., McCoy, D., Levchenko, K., Savage, S., and Voelker, G.~M.}
\newblock An analysis of underground forums.
\newblock In {\em ACM SIGCOMM Conference on Internet Measurement Conference (IMC)\/} (2011).

\bibitem{socialmediafraudFTC}
{\sc News, F.}
\newblock Ftc data shows consumers report losing \$2.7 billion to social media scams since 2021.
\newblock \url{https://www.ftc.gov/news-events/news/press-releases/2023/10/ftc-data-shows-consumers-report-losing-27-billion-social-media-scams-2021}, 2023.

\bibitem{influencefraud}
{\sc News, F.}
\newblock J'finfluencers’ charged for promoting unauthorised trading scheme.
\newblock \url{https://www.fca.org.uk/news/press-releases/finfluencers-charged-promoting-unauthorised-trading-scheme}, 2024.

\bibitem{sextortion}
{\sc News, W.~P.}
\newblock The rise of sextortion and responses to a growing crime.
\newblock \url{https://www.weprotect.org/issue/sextortion/}.

\bibitem{fraudjobs}
{\sc Popovici, M.}
\newblock Job scams report – 2,670 social media posts reveal scammers top tactics.
\newblock \url{https://heimdalsecurity.com/blog/job-scam-social-media-study/}, 2024.

\bibitem{shippingscams}
{\sc Puig, A.}
\newblock Fake shipping notification emails and text messages: What you need to know this holiday season.
\newblock \url{https://consumer.ftc.gov/consumer-alerts/2023/12/fake-shipping-notification-emails-and-text-messages-what-you-need-know-holiday-season}, 2023.

\bibitem{reimers2019sentencebert}
{\sc Reimers, N., and Gurevych, I.}
\newblock {Sentence-BERT: Sentence Embeddings using Siamese BERT-Networks}.
\newblock In {\em Empirical Methods in Natural Language Processing (EMNLP)\/} (2019).

\bibitem{Ruan2016Profiling}
{\sc Ruan, X., Wu, Z., Wang, H., and Jajodia, S.}
\newblock Profiling online social behaviors for compromised account detection.
\newblock {\em IEEE Transactions on Information Forensics and Security (ITIFS)\/} (2016).

\bibitem{socialmediaScamRise2}
{\sc Sebastain, N.}
\newblock Social media scams: Stunning statistics and tips to protect yourself.
\newblock \url{https://www.goodfirms.co/resources/social-media-scams-statistics-and-tips-for-protection}, 2024.

\bibitem{stivala2020deceptive}
{\sc Stivala, G., and Pellegrino, G.}
\newblock Deceptive previews: A study of the link preview trustworthiness in social platforms.

\bibitem{stringhini2012poultry}
{\sc Stringhini, G., Egele, M., Kruegel, C., and Vigna, G.}
\newblock Poultry markets: on the underground economy of twitter followers.
\newblock {\em ACM SIGCOMM Computer Communication Review\/} (2012).

\bibitem{Stringhini2010Detecting}
{\sc Stringhini, G., Kruegel, C., and Vigna, G.}
\newblock {Detecting Spammers on Social Networks}.
\newblock In {\em Annual Computer Security Applications Conference (ACSAC)\/} (2010).

\bibitem{stringhini2013follow}
{\sc Stringhini, G., Wang, G., Egele, M., Kruegel, C., Vigna, G., Zheng, H., and Zhao, B.~Y.}
\newblock Follow the green: growth and dynamics in twitter follower markets.
\newblock In {\em ACM SIGCOMM Conference on Internet Measurement Conference (IMC)\/} (2013).

\bibitem{researchcode}
{\sc SysSec}.
\newblock {Buy and Sale of Social Media Code and Data}.
\newblock \url{https://github.com/CISPA-SysSec/social_media_buy_and_sale}, 2024.

\bibitem{thomas2013trafficking}
{\sc Thomas, K., McCoy, D., Grier, C., Kolcz, A., and Paxson, V.}
\newblock $\{$Trafficking$\}$ fraudulent accounts: The role of the underground market in twitter spam and abuse.
\newblock In {\em USENIX Security\/} (2013).

\bibitem{traang2015evaluating}
{\sc Tr{\aa}ng, D., Johansson, F., and Rosell, M.}
\newblock Evaluating algorithms for detection of compromised social media user accounts.
\newblock In {\em 2015 Second European Network Intelligence Conference\/} (2015), European Network Intelligence Conference (ENIC).

\bibitem{TwitterUserDetailAPI}
{\sc Twitter}.
\newblock User detail twitter api.
\newblock \url{https://developer.twitter.com/en/docs/twitter-api/v1/accounts-and-users/follow-search-get-users/api-reference/get-users-lookup}, 2024.

\bibitem{TwitterTimelinesAPI}
{\sc Twitter}.
\newblock User timelines twitter api.
\newblock \url{https://developer.twitter.com/en/docs/twitter-api/tweets/timelines/introduction}, 2024.

\bibitem{viswanath2014towards}
{\sc Viswanath, B., Bashir, M.~A., Crovella, M., Guha, S., Gummadi, K.~P., Krishnamurthy, B., and Mislove, A.}
\newblock Towards detecting anomalous user behavior in online social networks.
\newblock In {\em Usenix Security\/} (2014).

\bibitem{socialmediafakeproducts}
{\sc Williams, R.}
\newblock The growth of fake products on social media.
\newblock \url{https://www.redpoints.com/blog/the-growth-of-fake-products-on-social-media/}, 2024.

\bibitem{xiao2015detecting}
{\sc Xiao, C., Freeman, D.~M., and Hwa, T.}
\newblock Detecting clusters of fake accounts in online social networks.
\newblock In {\em ACM Workshop on Artificial Intelligence and Security (AIS)\/} (2015).

\bibitem{xu2021deep}
{\sc Xu, T., Goossen, G., Cevahir, H.~K., Khodeir, S., Jin, Y., Li, F., Shan, S., Patel, S., Freeman, D., and Pearce, P.}
\newblock Deep entity classification: Abusive account detection for online social networks.
\newblock In {\em USENIX Security\/} (2021).

\end{thebibliography}


\appendix

\begin{table*}
  \centering
  \scriptsize
  \caption{Overview of trading channels identified. The table marks all the trading channels monitored in our study, with others not containing account handles publicly or being infeasible to be tracked due to crawling challenges such as CAPTCHAs, complex user interactions, and analysis prerequisites like account credentials.}

  \label{tab:appendix_whole_collected_data}
  \begin{tabular}{l|lll|ccc}

    \textbf{} &
    \multicolumn{3}{c|}{\faGlobe~\textbf{Exchange}} &
    \multicolumn{3}{c}{\faUser~\textbf{Accounts}} \\

    \toprule
    \textbf{Category} &
    \textbf{Channel} &
    \textbf{Type}     &
    \textbf{Source}  &
    \textbf{Selling} &
    \textbf{Handles}   &
    \textbf{Monitored} \\

    \midrule

            Public 

            & accs-market.com               & Marketplace    & Google Search                & \fullcircle & \fullcircle & \fullcircle   \\
            & fameswap.com                  & Marketplace    & Google Search                & \fullcircle & \fullcircle & \fullcircle   \\
            & www.z2u.com                   & Marketplace    & Google Search                & \fullcircle & \fullcircle & \fullcircle   \\
            & fameseller.com                & Marketplace    & Google Search                & \fullcircle & \fullcircle & \fullcircle  \\
            & insta-sale.comlistings/       & Marketplace    & Google Search                & \fullcircle & \fullcircle & \fullcircle  \\
            & accsmarket.com                 & Shop           & Google Search                & \fullcircle & \fullcircle & \fullcircle  \\
            & buysocia.com                  & Shop           & Google Search                & \fullcircle & \fullcircle & \fullcircle  \\
            & mid-man.com                   & Shop           & Google Search                & \fullcircle & \fullcircle & \fullcircle  \\
            & socialtradia.com              & Shop           & Google Search                & \fullcircle & \fullcircle & \fullcircle  \\
            & swapsocials.com               & Shop           & Google Search                & \fullcircle & \fullcircle & \fullcircle  \\
            & www.surgegram.com             & Shop           & Google Search                & \fullcircle & \fullcircle & \fullcircle  \\
            & www.toofame.com               & Shop           & Google Search                & \fullcircle & \fullcircle & \fullcircle  \\
            & cracked.io                     & Marketplace    & \cite{lykousas2023cynicism}  & \fullcircle & \emptycircle  & \fullcircle   \\
            & hackforums.net                 & BlackHat Forum & Google Search                & \fullcircle & \emptycircle  & \fullcircle   \\
            & swapd.co                      & Marketplace    & Google Search                & \fullcircle & \emptycircle  & \fullcircle   \\

            & accszone.com                   & Shop           & Public BH Forum              & \fullcircle & \emptycircle  & \emptycircle  \\
            & agedprofiles.com               & Shop           & Public BH Forum              & \fullcircle & \emptycircle  & \emptycircle  \\
            & bulkacc.com                   & Shop           & Public BH Forum              & \fullcircle & \emptycircle  & \emptycircle  \\
            & digitalchaining.mysellix.io    & Shop           & Public BH Forum              & \fullcircle & \emptycircle  & \emptycircle  \\
            & discord.gg/PMJCYxCcCu          & Shop           & Public BH Forum              & \fullcircle & \emptycircle  & \emptycircle  \\
            & nwarlordyt.sellpass.io         & Shop           & Public BH Forum              & \fullcircle & \emptycircle  & \emptycircle  \\
            & famousinfluencer.com           & Shop           & Public BH Forum              & \fullcircle & \emptycircle  & \emptycircle  \\
            & nloaccs.com                   & Shop           & Public BH Forum              & \fullcircle & \emptycircle  & \emptycircle  \\
            & www.smmzone24.com              & Shop           & Public BH Forum              & \fullcircle & \emptycircle  & \emptycircle  \\
            & acccluster.com                & Shop           & Google Search                & \fullcircle & \emptycircle  & \emptycircle  \\
            & accsmaster.com                & Shop           & Google Search                & \fullcircle & \emptycircle  & \emptycircle  \\
            & buyaccs.com                    & Shop           & \cite{thomas2013trafficking} & \fullcircle & \emptycircle  & \emptycircle  \\
            & getbulkaccounts.com            & Shop           &  \cite{thomas2013trafficking} & \fullcircle    & \emptycircle    & \emptycircle     \\
            & (bulkye.com)                   & Shop           & \cite{thomas2013trafficking} & \fullcircle & \emptycircle  & \emptycircle  \\
            & quickaccounts.bigcartel.com    & Shop           & \cite{thomas2013trafficking} & \fullcircle & \emptycircle  & \emptycircle  \\
            & twiends.com                   & BlackHat Forum & \cite{stringhini2013follow}  & \emptycircle  & \emptycircle  & \emptycircle  \\
            & leakzone.net /                 & BlackHat Forum & Google Search                & \emptycircle  & \emptycircle  & \emptycircle  \\
            & magicsmm.com                  & Shop           & Public BH Forum              & \emptycircle  & \emptycircle  & \emptycircle  \\
            & paneliniz.net                  & Shop           & Public BH Forum              & \emptycircle  & \emptycircle  & \emptycircle  \\
            & smmorigins.com                 & Shop           & Public BH Forum              & \emptycircle  & \emptycircle  & \emptycircle  \\
            & smmtake.com                    & Shop           & Public BH Forum              & \emptycircle  & \emptycircle  & \emptycircle  \\
            & bigfollow.net                 & Shop           & \cite{stringhini2013follow}  & \emptycircle  & \emptycircle  & \emptycircle  \\
            & intertwitter.com              & Shop           & \cite{stringhini2013follow}  & \emptycircle  & \emptycircle  & \emptycircle  \\
            & seguidores.com.br              & Shop           & Redirect from bigfollow      & \emptycircle  & \emptycircle  & \emptycircle  \\
            & scrowise.com                  & Shop           & Google Search                & \emptycircle  & \emptycircle  & \emptycircle  \\

    \midrule
    Underground

                & Dark Matter                    & Marketplace    & Onion Directory              & \fullcircle & \emptycircle  & \fullcircle  \\
                & Nexus Market                   & Marketplace    & Onion Directory              & \fullcircle & \emptycircle  & \fullcircle  \\
                & Torzon Market                  & Marketplace    & Onion Directory              & \fullcircle & \emptycircle  & \fullcircle  \\
                & Black Pyramid                  & Marketplace    & Onion Directory              & \fullcircle & \emptycircle  & \fullcircle  \\
                & Kerberos                       & Marketplace    & \cite{lin2024malla}          & \fullcircle & \emptycircle  & \fullcircle  \\
                & WeTherth                     & Marketplace    & \cite{lin2024malla}          & \fullcircle & \emptycircle  & \fullcircle  \\
                & MGM Grand                      & Marketplace    & \cite{lin2024malla}          & \fullcircle & \emptycircle  & \emptycircle  \\
                & ARES market                    & Marketplace    & Onion Directory              & \fullcircle & \emptycircle  & \emptycircle  \\
                & Soza                         & Marketplace    & Onion Directory              & \emptycircle  & \emptycircle  & \emptycircle  \\
                & SuperMarket                    & Marketplace    & Onion Directory              & \emptycircle  & \emptycircle  & \emptycircle  \\
                & Quantum Market                 & Marketplace    & Onion Directory              & \emptycircle  & \emptycircle  & \emptycircle  \\
                & Quest Market                   & Marketplace    & Onion Directory              & \emptycircle  & \emptycircle  & \emptycircle  \\
                & Incognito                      & Marketplace    & Onion Directory              & \emptycircle  & \emptycircle  & \emptycircle  \\
                & Alias Market                   & Marketplace    & Onion Directory              & \emptycircle  & \emptycircle  & \emptycircle  \\
                & Archetyp                       & Marketplace    & Onion Directory              & \emptycircle  & \emptycircle  & \emptycircle  \\
                & City Market                    & Marketplace    & Onion Directory              & \emptycircle  & \emptycircle  & \emptycircle  \\
                & Elysium                        & Marketplace    & Onion Directory              & \emptycircle  & \emptycircle  & \emptycircle  \\
                & Fish Market                    & Marketplace    & Onion Directory              & \emptycircle  & \emptycircle  & \emptycircle  \\
                & Pegasus Market                 & Marketplace    & Onion Directory              & \emptycircle  & \emptycircle  & \emptycircle  \\
                & Abacus                         & Marketplace    & \cite{lin2024malla}          & \emptycircle  & \emptycircle  & \emptycircle  \\

    \midrule
    Contact    & Skyisthelimitservice@gmail.com & Email          & Public BH Forum              & \fullcircle & \emptycircle  & \emptycircle  \\
               & t.me/BusinessAts               & Telegram       & Public BH Forum              & \fullcircle & \emptycircle  & \emptycircle  \\
               & t.me/sheriff\_x                & Telegram       & Public BH Forum              & \fullcircle & \emptycircle  & \emptycircle  \\
               & t.me/igexpertbhw               & Telegram       & Public BH Forum              & \fullcircle & \emptycircle  & \emptycircle  \\
               & t.me/lulpola                   & Telegram       & Public BH Forum              & \fullcircle & \emptycircle  & \emptycircle  \\
               & t.me/prudentagency11           & Telegram       & Public BH Forum              & \fullcircle & \emptycircle  & \emptycircle  \\
               & t.me/gunnupgrades              & Telegram       & Public BH Forum              & \fullcircle & \emptycircle  & \emptycircle  \\
               & +16193762832                   & Whatsapp       & Public BH Forum              & \fullcircle & \emptycircle  & \emptycircle  \\
               & @gunnupg                       & Discord        & Public BH Forum              & \fullcircle & \emptycircle  & \emptycircle  \\
               & @MaxRuslan369                  & Unknown        & Public BH Forum              & \fullcircle & \emptycircle  & \emptycircle  \\

    \bottomrule
  \end{tabular}
\end{table*}

\section{Public Marketplaces Payment Methods Additional Details}
\label{sec:payment_supported}

In this section, we analyze the supported payment methods by the marketplaces and their security implications. 

\subsection{Payment Method Extraction}

To identify the payment methods supported by each marketplace, we conducted a comprehensive manual analysis. For each marketplace, we visited its publicly available website and carefully navigated through relevant sections such as payment pages, FAQs, user guides, or checkout interfaces from multiple vantage points, as certain payment methods might only be visible or available to users accessing the platform from specific regions. This ensured that we gathered the most accurate and up-to-date information on payment methods without relying solely on indirect sources like Google search. We recorded all payment methods explicitly listed or implied by the marketplace, such as PayPal, cryptocurrencies, and alternative methods like WeChat Pay and Skrill.

We noted whether the payment methods were visible without requiring user interaction (e.g., creating an account or initiating a purchase). If details were not immediately visible, additional steps such as creating an account were undertaken when necessary. To ensure the accuracy of the collected data, we cross-verified each marketplace's payment methods with multiple pages on the marketplace. For platforms with unclear or incomplete information, we performed test interactions, such as attempting to initiate a test transaction, to confirm the availability of specific payment methods.

\subsection{Security Implications}

The analysis of supported payment methods across marketplaces reveals a significant variation in availability, reflecting differing priorities in terms of accessibility, user convenience, and security. Table~\ref{tab:payment_methods} presents the supported payment methods across different marketplaces. Overall, marketplaces that prioritize transparent payment methods and adopt systems with strong buyer protection, such as PayPal and Skrill, provide a safer environment for users. Conversely, reliance on cryptocurrencies or undisclosed payment options increases risks of fraud and dispute resolution challenges.

\BfPara{Risk of Irreversible Payments} We observed a wide support for Bitcoin (BTC), Ethereum (ETH), and other cryptocurrencies across marketplaces. While cryptocurrencies enable anonymous transactions, they introduce higher risks due to their irreversible nature and the potential for fraud or illicit activities without buyer protection mechanisms.

\BfPara{Buyer Protection and Chargebacks} Digital wallets such as PayPal and Skrill can reduce the exposure of bank card details, and offer users strong buyer protection, including refunds and chargebacks. However, these payment methods are adopted only by two marketplaces (Z2U and FameSeller). 

\BfPara{Regional Payment Methods and Vouchers} NeoSurf and Payssion, supported by select marketplaces like Z2U, cater to regional or prepaid needs, providing alternatives to bank-linked systems. These methods enhance user privacy by not linking transactions to bank accounts or personal information but offer limited recourse in disputes or fraud cases.

\BfPara{Escrow-Like Systems} Trustap and Payer, available on MidMan and TooFame, enhance security by holding funds in escrow until predefined conditions are met, reducing fraud risks for high-value or deferred-delivery transactions. However, their effectiveness depends on the trustworthiness and terms of the escrow provider.

\BfPara{Transparency of Payment Methods} For marketplaces such as Accsmarket, FameSwap, and TooFame, payment methods were \textit{unknown} or not publicly disclosed, increasing the likelihood of users interacting with unprotected or insecure systems.

\end{document}